# Characterization of Large Volume 3.5" x 8" LaBr$_3$:Ce Detectors

A. Giaz[a,b], L. Pellegri[a,b], S. Riboldi[a,b], F. Camera[a,b,**], N. Blasi[b], C. Boiano[b], A. Bracco[a,b], S. Brambilla[b], S. Ceruti[a], S. Coelli[b], F.C.L. Crespi[a,b], M. Csatlòs[c], S. Frega[a], J. Gulyàs[c], A. Krasznahorkay[c], S. Lodetti[a], B. Million[b], A. Owens[d], F. Quarati[d,*], L. Stuhl[c], and O. Wieland[b]

[a] Università degli Studi di Milano, Physics Dept., Via Celoria 16, 20133 Milano, Italy
[b] INFN Milano, Via Celoria 16, 20133 Milano, Italy
[c] Institute of Nuclear Research of the Hungarian Academy of Sciences (ATOMKI), P.O. Box 51, H-4001, Debrecen, Hungary
[d] SRE-PA, ESA/ESTEC, Keplerlaan 1, 2200AG Noordwijk, The Netherlands

*Present address Delft University of Technology, I.R.I., Mekelweg 15, 2629 JB Delft, The Netherlands

**Corresponding author: Franco Camera (Franco.Camera@mi.infn.it)

**Abstract**

The properties of large volume cylindrical 3.5" x 8″ (89 mm x 203 mm) LaBr$_3$:Ce scintillation detectors coupled to the Hamamatsu R10233-100SEL photo-multiplier tube were investigated. These crystals are among the largest ones ever produced and still need to be fully characterized to determine how these detectors can be utilized and in which applications. We tested the detectors using monochromatic γ–ray sources and in-beam reactions producing γ rays up to 22.6 MeV; we acquired PMT signal pulses and calculated detector energy resolution and response linearity as a function of γ-ray energy. Two different voltage dividers were coupled to the Hamamatsu R10233-100SEL PMT: the Hamamatsu E1198-26, based on straightforward resistive network design, and the "LABRVD", specifically designed for our large volume LaBr$_3$:Ce scintillation detectors, which also includes active semiconductor devices. Because of the extremely high light yield of LaBr$_3$:Ce crystals we observed that, depending on the choice of PMT, voltage divider and applied voltage, some significant deviation from the ideally proportional response of the detector and some pulse shape deformation appear. In addition, crystal non-homogeneities and PMT gain drifts affect the (measured) energy resolution especially in case of high-energy γ rays. We also measured the time resolution of detectors with different sizes (from 1"x1" up to 3.5"x8"), correlating the results with both the intrinsic properties of PMTs and GEANT simulations of the scintillation light collection process. The detector absolute full energy efficiency was measured and simulated up to γ-rays of 30 MeV

## 1. Introduction

The Cerium doped Lanthanum bromide material is an inorganic scintillator, made available to the scientific community only a few years ago [1-2]. It has excellent properties, e.g. the best energy resolution among all the scintillators (2.7-3.3% FWHM at 661.6 keV), sub-nanosecond time resolution, almost perfect light yield proportionality (down to about 100 keV) and good stability of the emitted light with temperature [1-14]. Material density is relatively high: 5.1 g/cm$^3$, to be compared with NaI:Tl (3.67 g/cm$^3$), BGO (7.13 g/cm$^3$) and HPGe (5.32 g/cm$^3$). However, crystals are both extremely fragile and highly hygroscopic, so that they must be kept and operated in sealed capsules. A detailed study of the scintillation signals has also shown the possibility to discriminate between α particles and γ rays by means of dedicated pulse shape analysis techniques [15-17].

The excellent properties of LaBr$_3$:Ce scintillation detectors have generated a large interest in the scientific community. This new material not only promises to be the best scintillation crystal for γ–ray detection/spectroscopy, but it can also be a possible, simpler and/or cheaper alternative to High Purity Germanium (HPGe) detectors. With this in mind, a LaBr$_3$:Ce-based detector array (possibly coupled with HPGe detectors) could operate as an extremely efficient, cost-effective and easy to use setup for γ-ray experiments. The effectiveness of scintillators with respect to HPGe detectors could be especially evident in case the γ-ray spectra to be measured are not very complex in nature or in case where the energy broadening of γ rays caused by the Doppler effect is larger than the intrinsic resolution of HPGe detectors [18-20].

Thanks to very good intrinsic time (< 1 ns), energy resolution and good detection efficiency of high-energy γ rays, large volume $LaBr_3$:Ce detectors can in principle provide at the same time clean spectroscopic information from a few tens of keV up to tens of MeV, being furthermore able to clearly separate the full energy peak from the first escape peak up to at least a γ-ray energy of 25 MeV. This is the case, for example, of specific in-beam γ–spectroscopy experiments with fast, exotic beams. Moreover, the excellent timing properties of $LaBr_3$:Ce scintillators also allow the acquisition of high resolution time information and to implement effective neutron–γ discrimination and background rejection, a critical point in the case of experiments with radioactive beams. Extremely intense, polarized and almost monochromatic γ-ray beams in the energy range between 1 MeV and 25 MeV will be readily available, in fore coming facilities [21-24]. In such facilities it will be possible, for example, to study highly collective nuclear states like the Giant or the Pygmy Dipole Resonance [25-31] by means of Nuclear Resonance Fluorescence (NRF) using high-energy γ rays as incident beam. Being able to efficiently identify high-energy γ rays, Lanthanum bromide detectors could thus additionally enforce the Physics program of HPGe detector arrays.

While the first small $LaBr_3$:Ce crystals were grown around 2001, only in 2008 the crystal manufacturer, Saint-Gobain Crystals, [1], was able to grow and distribute large volume 3.5"x8" crystals. Because of efficiency considerations, the availability of large volume crystals is a key aspect in the design of a high-energy γ-ray detector array. Quite a large amount of works with small sized $LaBr_3$:Ce detectors can be found in the literature [7, 8, 12, 14, 19], but only very few works related to medium volume detectors are available [7, 13, 32-39] and even less information is available for large volume $LaBr_3$:Ce detectors [40].

It is also important to point out that the properties of large volume LaBr3:Ce crystals cannot be easily derived from those of small and medium sized detectors. In fact, several factors may affect the detector performances: i) self absorption, ii) possible crystal internal non-homogeneities that may result in variation of the crystal light yield depending on the detector area affected by the interacting γ ray (both of which are more likely to appear with scaled up dimensions), iii) the much longer mean free path of the scintillation light towards the photo-cathode and iv) non-ideal photo-multiplier tube (PMT) properties. It is worth mentioning, for example, that the 50% absorption length in $LaBr_3$:Ce changes from 15 mm in case of 500 keV γ rays to 40 mm in case of 5 MeV γ rays [1].

In order to fully characterize the energy resolution performance of our large volume $LaBr_3$:Ce detectors, together with the associated PMT, voltage divider and subsequent electronics as a function of the measured energy, we acquired γ rays with energy spanning over three orders or magnitude.

In Section 2 we will discuss the pulse shape of the PMT output signal and how it can be affected by detector size, choice of the PMT and applied voltage. We have also inserted a short discussion concerning the mean free path of the scintillation light as a function of detector size. In Section 3 we will present the results of the detector response for various count rates of events (5-250 kHz). In Section 4 we will discuss large volume 3.5"x8" $LaBr_3$:Ce detector response to γ rays in the energy range between 5 keV and 22.6 MeV, specifically analyzing, in Section 4.2, the pulse shape of the PMT output signal as a function of γ-ray energy and the choice of voltage divider. In Section 4.3 we will discuss the detector behavior in terms of linearity of response, in Section 4.4, the corresponding energy resolution and in Section 4.5 the estimated detector time response. In Section 5 we will finally compare the experimentally measured efficiency of the large volume $LaBr_3$:Ce scintillation detectors against the results of GEANT3 simulations.

## 2. LaBr$_3$:Ce pulse signals

It is well known from the literature that the intrinsic time distribution of photon emission in LaBr$_3$:Ce scintillators shows an almost instantaneous rise following the γ-ray interaction, and a subsequent exponential decay with time constant of about 16 ns [1]. However, this is not the case for our crystals, as the electrical signals measured at the PMT anode show much slower rise and fall times.

We separately tested several LaBr$_3$:Ce cylindrical crystals (varying in size from 1"x1", 25 mm x 25 mm, to 3"x3", 76 mm x 76 mm, up to 3.5"x8", 89 mm x 203 mm) coupled to several PMTs of different area, operated at various high voltage levels and acquired the corresponding signal pulse shapes.

Table 1 summarizes the configuration of the tested LaBr$_3$:Ce scintillation detectors, the corresponding area-matched PMTs and their associated voltage dividers. The nominal operating voltage level for all the PMTs, depending also on the choice of the voltage divider, is in the range between 500 V and 1000 V. In addition to the commercially available voltage dividers (model 184K/T, model AS20 by Saint-Gobain Crystals and model E1198-26 by Hamamatsu) we also used a custom made voltage divider identified in the table as "LABRVD", which was specifically designed at the University of Milano for our LaBr$_3$:Ce crystals [41]. It mainly consists of: i) a resistive divider network that sets the bias voltage levels of the PMT dynodes; ii) a subsequent P-channel MOSFET network that operates as a high impedance voltage buffer and iii) a last PNP BJT network that provides voltage buffering with high current capability. Reliable operation of the PMT voltage divider is ensured by protecting diodes, even in case of abrupt changes in the high voltage level, e.g. as a consequence of unexpected disconnection of cabling. The voltage divider ratios have been experimentally tuned to preserve the intrinsically good crystal properties in terms of energy and time measurements, while obtaining at the same time relatively homogeneous performance in terms of energy response linearity among the various PMT parts. Namely, the voltage difference between the photo-cathode and the first dynode was increased by 50% with respect to the average value among dynodes in order to improve the timing properties, while the potential of the very last dynode and the voltage difference between the last two dynodes were respectively increased by 30% and 50% in order to improve the linearity of PMT energy response. Individual tuning of the voltage divider ratios for each PMT could in principle provide even better detector performance, especially in terms of linearity of energy response, nonetheless we chose to operate all the PMTs by just using a standard VD model, mainly to preserve interchangeability and reduce the need of dedicated spare parts.

Table 1 also quotes the intrinsic rise-time of the single photon response of the various PMTs with the associated voltage divider operated at nominal high voltage level, either experimentally measured or derived from the manufacturer datasheet. The reference [8] discusses the performances of the non-standard Photonis XP20D0B mounted on the sealed detector 1 in table 1. Not surprisingly, as a general rule, large area spectroscopic PMTs show intrinsic slow single photon response rise-time, thus substantially contributing to the overall increase of the rise and fall times of our large volume (3.5"x8") LaBr$_3$:Ce detectors.

**Table 1**: *The six detector configurations used to investigate signal pulse shapes of our LaBr$_3$:Ce detectors, summarized in terms of crystal size, associated area-matched PMT and corresponding voltage divider. The intrinsic rise time of the PMT response to single photon is also shown.*

**Figure 1**. *The rise-time (10% - 90%) of the anode signal of the six LaBr$_3$:Ce detector configurations listed in table 1 (estimated with 1 ns uncertainty) in case of 661.6 keV energy pulses, as a function of the high voltage level supplied to the PMT voltage divider (color online). The numbers in the legend correspond to those of table1.*

Figure 1 shows the rise time (10% - 90%) of the six LaBr$_3$:Ce detector configurations listed in table 1 (estimated with 1 ns uncertainty) in case of 661.6 keV energy pulses, acquired using a 600

MHz bandwidth, 2 GHz sampling frequency digital oscilloscope (Lecroy Waverunner HRO66Zi). For any given detector configuration, increasing the high voltage level supplied to the PMT voltage divider (up to reasonable values depending on the PMT and voltage divider itself) always leads to faster PMT output signals. The minimum rise-time practically obtainable with the six configurations of table 1 ranges from 6 ns up to 23 ns. By comparing the experimental rise-time of the six LaBr$_3$:Ce detectors in figure 1 with the single photon emission rise-time of the corresponding PMT in table 1, there is a clear evidence that PMT properties alone cannot be the only cause of signal slowdown.

In order to better understand our experimental results, namely to disentangle the effect of the increased collection time of the scintillation light from the effect induced by a large surface PMT, we separately coupled three LaBr$_3$:Ce crystals (1"x1", 3"x3" and 3.5"x8" in size) to a single, small and fast PMT (model H6533 by Hamamatsu), operated with the incorporated voltage divider. This PMT, operated with nominal high voltage level, has a single photon response rise-time of less than 0.7 ns, so that it can be considered fast enough to almost completely preserve the time properties of LaBr$_3$:Ce pulses.

For each of the three detector configurations we acquired a few tens of 661.6 keV energy pulses using a 400 MHz bandwidth, 5 GHz sampling frequency digital oscilloscope (LeCroy Waverunner 44X1) and subsequently extracted the three average signals shown in figure 2 with solid lines. In fact, the averaging of the pulses of the corresponding data sets allows to improve the signal to noise ratio. This is important especially in the case of the last two crystals for which only a small fraction of the emitted photons were actually collected by the H6533 PMT. The three signals (a) of figure 2 reflects the time required to collect the scintillation light and should change with the detector size. Clearly, both the rise and the decay-time pulses slows down with increased crystal size, varying from 4 ns (1"x1" crystal) to 7 ns (3"x3" crystal) and up to 14 ns (3.5"x8" crystal). This direct correlation between detector size and pulse rise time, given a fixed PMT assembly, may only be the result of a larger spread in the collection time of the LaBr$_3$:Ce-emitted photons at the PMT entrance window, as a consequence of longer light paths towards the photo-cathode.

These results are consistent with the experimentally measured rise times of figure 1, once the specific time properties of the five PMTs are taken into account. Indeed, by calculating the convolution product of the three crystal-dependent reference signals of figure 2 (a) with the single photon time response of the associated PMTs (see table 1) we obtained the pulses (b) indicated by dashed lines. The rise times resulting from the convolution product (7 ns for 1"x1" crystal, 19 ns for 3"x3" one and 23 ns for 3.5"x8" one) are very well in agreement with the measured anode signals of the detector assemblies, as shown in figure 1.

All the previously reported experimental results are also in agreement with the collection process of optical photons inside LaBr$_3$:Ce detectors described in the literature [13], simulated using GEANT4 and the 'unified model' libraries. These simulations show, for example, that the average path length of the scintillation photons in case of a 4x4x5 mm$^3$ LaBr$_3$:Ce crystal is 31 mm, increasing to 390 mm in case of a 51 mm x 76 mm cylindrical crystal. The GEANT4 simulations that we performed for our even larger crystals predicted an average path length of 450 mm for the 76x76 mm (3"x3") crystal and of 1200 mm for the 89x203 mm (3.5"x8") one. Such large values are direct consequences of two effects: on one hand the average path length increases with the detector length and on the other hand, the very different refractive index between the LaBr$_3$:Ce crystal (n ≈ 1.9) and borosilicate glass (n ≈ 1.5) makes the large majority of the scintillation photons undergo a many-times reflection process before being absorbed by the PMT photo-cathode.

**Figure 2**: *The LaBr3:Ce signal pulse shapes (solid lines), indicated by (a), obtained with the Hamamatsu H6533 PMT coupled to our three cylindrical crystals (1"x1", 3"x3" and 3.5"x 8"). The difference in the pulse shapes is a direct consequence of the different spread in the collection time of the scintillation photons. The signals (dashed-lines) indicated by (b), are the convolution product of the three crystal signals, measured with the Hamamatsu H6533 PMT, (indicated by (a)) with the single photon time responses of the associated PMTs (color online).*

As a direct consequence of the slower rise and fall-times of LaBr$_3$:Ce pulses, it is evident that the estimation of the γ-ray interaction time will worsen in case of large volume crystals. However, as it will be shown in Section 4.5, we were still able to obtain a sub-nanosecond intrinsic time resolution. On the other hand, we expect that the performances of pulse shape analysis algorithms, like the one discussed in [16], could in general be somehow affected by the increased size of the detector. In fact, the difference between alpha particle and γ ray induced pulses discussed in ref. [16] will be reduced by the PMT longer intrinsic signal rise time [42].

## 3. Detector gain stability

As a remarkably positive feature, LaBr$_3$:Ce detector pulses have quite short time extension. In case of our large volume detectors, typical PMT anode signals induced by γ-ray interactions last for approximately 150 ns. Such fast pulses should in principle allow detectors to stand extremely high count rates of events [7,43-46]. However, this particular working condition can severely impact PMT operation and therefore degrade the overall detector performance. When the event count rate increases, both the average current in the PMT voltage divider and the flux of electrons inside the PMT increase. These phenomena produce opposite effects on the PMT conversion gain: while the latter effect slightly decreases the PMT gain, the former one, typically more evident, tends to increase the PMT gain.

Passive voltage dividers usually suffer the most from the latter effect. Because of the high impedance values at the PMT dynodes, large current signals may significantly reduce the voltage difference between the last dynode and the anode of the PMT, thus upsetting the overall distribution of the high voltage power supply among all the PMT dynodes and increasing the PMT gain [47].

The "LABRVD" active voltage divider described in the previous section has been specifically developed to overcome this problem. We then tested the behavior of our large volume LaBr3:Ce detectors in case of high count rate of events, with both a commercially available voltage divider and our custom-made solution in order to verify the effectiveness of the custom voltage divider.

This is a key issue to investigate in case of large volume detectors, since their higher γ-ray detection efficiency and larger solid angle give rise to high count rate, with consequent large average anode operational current in the associated PMT's. In addition, during in-beam experiments, detectors may also be subject to non negligible fluctuations of the beam intensity over time, leading to correlated gain changes in the event count rate. In case of high-energy γ rays and no dedicated off-line correction, the combined effect of even relatively small PMT gain drifts (e.g. of the order of 0.5%) can significantly deteriorate the overall detector energy resolution (e.g. intrinsically around 0.85% FWHM at 9 MeV).

We acquired and processed LaBr$_3$:Ce scintillation pulses from γ-ray sources up to 2 MeV using the last two detector configurations reported in table 1 (cases 4 and 5). The two corresponding voltage dividers ("LABRVD" in case 4 and model E1198-26 in case 5) were operated at the high voltage levels adjusted to obtain an anode pulse amplitudes of 30 mV, 60 mV and 90 mV with 50 Ohm load for 661.6 keV of deposited energy for both cases as listed in table 2. This approach makes identical the dynode current. As the gain of the PMTs could change significantly (even though the model is the same), it should be noticed that the high voltage levels reported in table 2 specifically refer to the PMT unit used in the test. The different values of HV to achieve the same gain is due to the different partition the two voltage dividers provide.

In order to estimate the count rate effect on both the energy resolution and the centroid position of a reference γ-line, two sources were used: a low activity $^{88}$Y source, placed at fixed distance from the detector, acts as the reference γ-line  and a 400 MBq $^{137}$Cs source was used to produce increasingly higher event count rates (from a few kHz up to 250 kHz) by being moved closer and closer towards the detector. We selected the $^{88}$Y source because its 898 keV line does not interfere with the 1323.2 keV sum peak produced by two concurrent 661.6 γ rays of the $^{137}$Cs.  We acquired PMT anode signals using a 400 MHz, 5 GHz sampling frequency digital oscilloscope (LeCroy Waverunner 44X1) with trigger threshold set above the 661.6 keV line level. Energy estimation was performed using a straightforward box-car integration algorithm (the sum of each digitized sample over 250 ns) with the additional subtraction of the pulse baseline level (calculated over 250 ns).

**Table 2:** *The high voltage levels applied to a single R10233-100SEL PMT unit coupled to two different voltage dividers ("LABRVD"- case 4 and E1198-26 - case 5; see table 1) in order to obtain PMT anode signals (with 50 Ohm load) of 30 mV, 60 mV and 90 mV in case of a deposited energy of 661.6 keV.*

Figures 3 and 4 show the extracted values of the centroid drifts and the FWHM energy resolution for the 898 keV $^{88}$Y peak, measured with a 3.5" x 8" LaBr$_3$:Ce detector operated at event count rate ranging from a few kHz up to 250 kHz. The top panel of figure 3 shows the values of the energy resolution obtained with the "LABRVD" active voltage divider (case 4 of table 1), while the bottom panel of figure 3 shows the resolution obtained with the Hamamatsu E1198-26 voltage divider (case 5 of table 1). As it can be observed, the two configurations give similar results provided the detector count rate is stable during all the measurement time. On the other hand, the two experimental configurations provide very different results for the centroid position of the 898 keV peak as a function of event count rate (see figure 4). As expected, the PMT gain clearly increases using the E1198-26 passive voltage divider, while the use of "LABRVD" active voltage divider allows the PMT to stand much higher event count rates without high gain variation. As an example, the drift induced on the 898 keV peak by PMT gain variations (from a pulse height nominal value of ≈ 135 mV/MeV) and count rate ranging from 5 to 250 kHz is of the order of 40 keV with the passive voltage divider (operated at 740 V) and only 5 keV with the active voltage divider (operated at 970 V).

The experimental data in the top panel of figure 4, acquired with the "LABRVD" active voltage divider, show nonetheless a small PMT gain reduction as the count rate increases. This second order effect cannot be simply explained by the voltage unbalancing at the PMT dynodes as in case of passive voltage dividers, because in that case the gain drift would be positive (over-linearity effect). Negative gain drifts are usually instead associated with the electrons induced shielding of the dynodes potentials inside the PMT [47]. However, according to this last hypothesis, PMT gain variations would immediately follow in time the associated variation of the event count rate, which is not our case as gain drifts accumulate slowly and progressively, reaching a stable plateau after several minutes after each count rate variation. We then expect that the underlying cause is related to temperature change in the PMT core, as a result of dynodes current variation which is related to the count rate. It is well known from the literature that increasing PMT temperature results in decreasing the PMT gain (approximately - 0.3% for one degree) [47]. In order to check this we measured the PMT gain at 5 kHz in two different situations: i) at the end of a long acquisition at the same event count rate (5 kHz) and ii) right after a long acquisition at 250 kHz. We found a remarkable difference in the two experimental values: in the second case, although the measured rate is only 5 kHz, the PMT gain is much closer to the one measured with a rate of 250 kHz. This is most probably due to the heating of the PMT because of the long acquisition at 250 kHz. According to the PMT gain-to-temperature dependence quoted in the manufacturer datasheet, the gain drifts at 250 kHz reported in the top panel of figure 4 would correspond to ≈ 1 degree temperature change in the PMT core, well compatible with even little current changes in the voltage divider.

**Figure 3**: *The FWHM energy resolution of the 898 keV $^{88}$Y γ–line at various detector counting rates, ranging from 5 to 250 kHz (color online). The top panel shows the results obtained with the "LABRVD" active voltage divider (case 4 of table 1) while in the bottom panel the E1198-26 voltage divider by Hamamatsu was used as the reference configuration (case 5 of table 1). The error bars are smaller than marker size.*

It is important to point out that reliable, reproducible results can only be obtained with high voltage units providing sufficient stability over time and with enough line and load regulation to guarantee precision of operation within the order of 0.1 V over the whole range of the expected event count rate. As some of the high voltage units under test were not able to guarantee such precise operation, it is important to verify the specifications in case high count rates of events are expected.

**Figure 4**: *The centroid drift, defined as the difference between the measured γ-ray energy and the real one, observed at 898 keV measured from 5 to 250 kHz (color in the online version). The top panel values were obtained with the "LABRVD" active voltage divider (case 4 of table 1) while the bottom panel values were instead obtained with the reference voltage divider (case 5 of table 1). The error bars are smaller than marker size.*

As a second major point, in order to acquire the experimental data, shown in figures 3 and 4, we took care of waiting for several minutes after varying the detector count rate, so that the PMT operating condition was always stable. Unfortunately, this is not the case for in-beam measurements, because of beam intensity fluctuations, the PMT operating condition may continuously change in a time-scale of seconds. As a consequence, significant degradation in the measured energy resolution could arise in case the detectors are not robust enough with respect to the effects of count rate fluctuations. Such kind of gain drifts could in principle be monitored using a tagged light emitting diode (LED) source operated inside the detector housing (as was for example done in the $BaF_2$ detectors of the HECTOR array [48]), at the cost of additional complexity in the experimental setup.

As a final note, it is also important to point out that not only the PMTs but also the subsequent electronics, e.g. shaping amplifier, analog to digital converter, etc. may easily impair $LaBr_3$:Ce detector performances, especially in case of high count rate of events and with lack of pile-up rejection circuits.

As an example, figure 5 shows the energy spectra of two sources, $^{60}$Co and $^{88}$Y, acquired using a 3.5" x 8" $LaBr_3$:Ce detector (case 4 of table 1), the BaFPRO [49-50] amplifier (700 ns bipolar shaping time) and the CAEN V785 ADC. The two energy spectra were acquired at detector count rate of 5 kHz (top panel) and 150 kHz (bottom panel). An additional $^{137}$Cs source at 661.6 keV, below the acquisition threshold, was used to increase the event count rate in the bottom panel spectrum. By comparing the two energy spectra it is evident that significant deteriorations due to pulse pile-up effects are present in the bottom panel spectrum.

To summarize, large volume $LaBr_3$:Ce detectors themselves can stand very high count rate of events with no performance degradation (at least up to 250 kHz); however, in case of significant fluctuation of the event count rates over time, active voltage dividers such as the "LABRVD" model may considerably improve PMT gain stability and thus preserve detector performance. Additional effects related to the detector count rate of events may be introduced by the subsequent electronics, if pile-up rejection circuits or algorithms are not implemented.

## 4. Detector response to γ rays

Several research papers in the literature show that the $LaBr_3$:Ce response in terms of number of scintillation photons emitted with respect to the energy deposited by γ rays is extremely linear [51-53], except in the very low-energy range below 100 keV [52]. Unfortunately, the very fast and intense flashes of $LaBr_3$:Ce scintillation light may easily produce non linear response in the PMTs [37, 39, 47, 54, 55] especially in case of events which deposit several MeV of energy. The photo-electron current peak induced on PMTs coupled to $LaBr_3$:Ce scintillators is indeed ≈ 25 times higher with respect to NaI(Tl) scintillators.

**Figure 5:** *Energy spectra of $^{60}$Co and $^{88}$Y sources acquired with a 3.5" x 8" $LaBr_3$:Ce detector (case 4 of table 1), the BaFPRO amplifier (700 ns bipolar shaping time) and the CAEN V785 ADC. The two energy spectra were acquired at detector count rate of 5 kHz (top panel) and 150 kHz (bottom panel). A $^{137}$Cs source (below trigger threshold) was used to increase the event count rate.*

In the following subsections we will discuss the properties of large volume 3.5" x 8" LaBr$_3$:Ce detectors in terms of signal pulse shape, linearity and time response to γ rays not only in the low to middle energy range, but also in the high-energy range up to 22.6 MeV, as such large detectors are very likely operated with high-energy γ rays.

**4.1 Radiation sources**

Testing LaBr$_3$:Ce detector response to γ rays in a wide energy range (from 5 keV up to 23 MeV) is not a trivial task.
In the energy range up to 2 MeV we used standard calibration sources, e.g. $^{60}$Co, $^{133}$Ba, $^{137}$Cs, $^{152}$Eu and $^{88}$Y, while for the energy range between 2 and 9 MeV we produced γ rays by coupling an AmBe source in a paraffin housing to natural Nickel (Am-Be-Ni) [56] as the radiative capture of thermal neutrons in natural Nickel produces several γ-ray emissions, with the strongest and highest in energy is at 8.997 MeV.
Monochromatic γ rays with energy above 9 MeV can be obtained only with accelerator-driven nuclear reactions, e.g. those reported in table 3, that were obtained at the Institute of Nuclear Research of the Hungarian Academy of Sciences (ATOMKI). More details about the facility, the reactions and the associated targets can be found in [32]. All the reactions except the last one were performed with the ATOMKI Van de Graaff accelerator with a beam intensity of the order of 2-3 µA , while the last one was performed with the ATOMKI cyclotron.

**Table 3:** *In the first column we report the reactions obtained at the ATOMKI Institute, in the second column the corresponding proton energy and in the third column the energy of the γ rays produced. The first 6 reactions were obtained with a Van de Graff accelerator, while the last one with a cyclotron.*

**4.2 Pulse distortion**

A direct evidence of non linear operation of the PMT and voltage divider assembly is the distortion of pulse shapes [57]. Pulse shape should in principle be independent of the amount of energy deposited in LaBr$_3$:Ce detectors, apart from a slight difference between interactions with γ rays or α particles [15-17]. Pulse shapes associated to γ–ray interactions of different energy (from 1 MeV up to 17 MeV) in a 3.5" x 8" LaBr$_3$:Ce detector coupled to the "LABRVD" active voltage divider (case 4 of table 1) are shown in figure 6. We set the PMT gain as to obtain 30 mV amplitude pulses for a deposited energy of 661.6 keV (with 50 Ohm load) and acquired the anode pulses with a 12 bit, 2 GHz sampling frequency digitizer (CAEN V1729). In order to increase the signal-to-noise ratio, we acquired and averaged over a few tens of detector pulses obtaining the signals of figure 6, normalized to unitary area.
As figure 6 shows, the amount of pulse distortion, although being always relatively small, tends to increase with increasing γ–ray energy. The basic parameters of the reference pulse shapes in figure 6, i.e. rise-time, fall-time and FWHM-time are summarized in figure 7 as a function of the deposited energy. This whole procedure was applied also in the case of a 3.5" x 8" LaBr$_3$:Ce detector coupled to the E1198-26 voltage divider (case 5 of table 1), obtaining the results shown in figures 8 and 9.

**Figure 6**: *Top panel: The normalized difference between two pulses, one at 17, 11, 9 and 6.1 MeV and a reference one at 1.173 MeV, namely (S(Eγ(1173))-S(Eγ(X)))/MAX(S(Eγ(1173))). Bottom Panel: The pulse shapes of a 3.5" x 8" LaBr$_3$:Ce detector normalized to unitary area (color in the online version,) obtained with the "LABRVD" active voltage divider and γ-ray energies in the range from 1 to 17 MeV (see legend).*

**Figure 7**: *The basic pulse shape parameters (color in the online version) extracted from the signals of figure 6. Lines are only for visual support. The error bars are smaller than marker size.*

**Figure 8**: *The pulse shapes of a 3.5" x 8" LaBr$_3$:Ce detector, normalized to unitary area (color in the online version). acquired with the E1198-26 voltage divider, for γ-ray energy in the range from 661 keV to 9 MeV. In the inset it is displayed the enlargement of the pulse lineshape in the interval between 50- 100 ns.*

A difference of approximately 5% in the pulse lineshape is not observed before the 17 MeV pulses with the LABRVD while with the E1198-26, the same difference is observed already for the 9 MeV pulse lineshape. As a consequence, the LABRVD voltage divider gives a better linearity in the detector energy response, as will be shown in section 4.3. Note that the good linearity up to 17 MeV, obtained with the LABRVD, will allow to operate the PMT in its best performance regime, without reducing the number of used dynodes in the PMT or lowering the operating Voltage when detecting high-energy γ rays, as should instead be done with the E1198-26 voltage divider to resume an acceptable linearity for energies higher than 9 MeV.

**Figure 9**: *The basic pulse shape parameters (color in the online version) extracted from the signals of figure 8. Dashed lines are only for visual support. The error bars are smaller than marker size.*

### 4.3 Energy linearity

Photo-multiplier tubes are usually employed to detect very low intensity light pulses, so that their nominal gain (i.e. the photo-electrons multiplying factor) is of the order of $10^5$-$10^6$ in case of standard voltage dividers supplied at nominal high voltage level. On the other hand, radiation detectors based on high light yield scintillators emit such a large amounts of photons (e.g. 63 photons/keV in case of LaBr$_3$:Ce crystals) that the associated PMTs, if operated at their nominal gain values would produce too high current signals at the anode output. It is thus not surprising that, in case the standard voltage dividers are used with the PMTs, scintillation detector manufacturers usually suggest to operate PMTs at much lower voltage levels with respect to the suggested nominal values in the PMT datasheet.

Although such an approach is generally valid for detection of low-to-medium energy γ rays, it can be not always satisfactory in case of medium-to-high energy, as the PMT output pulses could be high enough to saturate the electronics or at least introduce significant non linear effects in the PMT operation.

In such cases, the simplest and straightforward solution would be to extract the energy information from one of the last PMT dynodes (getting the time information from either the dynode or the anode) or to reduce the PMT gain even more, by further reducing the applied high voltage level. However, the extraction of a current signal from one of the last dynodes could produce inconsistencies on the PMT anode signal. In order to guarantee consistent behavior of the PMT in terms of linearity, gain, etc. with various counting rates, all dynodes potentials should be kept at relatively constant levels that do not depend on the instantaneous current drawn by the PMT. Providing the last dynodes with high impedance resistive biasing and reading-out pulses with ac-coupled low impedance loads may alter the dynode potentials and thus add significant artifacts on the shape of the pulses at the anode level, even of the order of a few percent. On the other side, operating PMTs at quite lower voltage than the nominal value, may significantly affect both energy and time resolution. For instance, in case the whole PMT gain is evenly distributed among the dynode stages, its associated variance may be calculated as $\sigma^2 = 1/(\delta-1)$ [58], where δ is the gain provided by each dynode stage. Also in case the PMT gain is not evenly distributed among all the dynode stages (a more realistic situation), the previous assumption still basically holds, so that the lower is the high voltage level, the higher is the amplification variance.

Moreover, as already discussed in Section 2, also the rise and fall-times of LaBr$_3$:Ce detector signals increase with decreased high voltage, so that the intrinsic detector time resolution is more easily achieved with higher voltages.

In case LaBr$_3$:Ce detectors are used for measurements in the low-to-medium energy range (up to 3 MeV) covered by standard calibration sources, all the PMT non-idealities reported in the previous sections, e.g. non linear response, gain drifts, etc. might be not so critical issues, provided some accurate detector calibration with second or even higher order polynomials is routinely performed. However, in a situation where also high-energy γ rays need to be detected [27, 29, 30, 31] and larger dynamic range (e.g. up to 30 MeV) is required, any detector non-linearity might easily become a critical issue, which is the case, for example, in the case of our 3.5" x 8" large volume LaBr$_3$:Ce detectors.

Complete characterization of the energy response linearity in case of large volume LaBr$_3$:Ce detectors should be performed using not only the standard calibration sources, but also all the γ rays discussed in section 4.1. However, such a procedure can be very time consuming and requires dedicated accelerator facilities, so that it cannot be easily repeated during standard experimental conditions. Up to now, we were able to individually characterize the energy response linearity of two large volume LaBr$_3$:Ce detectors. In order to quantify the linearity error in the calibrated energy response of the detector we first derived a linear calibration using only a few sources up to 2 MeV, using a PMT gain corresponding to 30 mV pulse amplitude for 661.6 keV γ rays, and then calculated the residuals between the real energy and the expected energy evaluated using the linear calibration. The calculated residuals are shown in figure 10, for both cases 4 and 5 of table 1 (LABRVD and E1198-26).

As figure 10 shows, we obtained good linearity in the detector energy response with the "LABRVD" voltage divider up to 17.6 MeV (we measure a relative deviation of less than 0.6%), with a steep increase above that energy (2.7% at 22.6 MeV), compared to the one obtained with the E1198-26 voltage divider which is above 1% already at 9 MeV. Such a good linearity implies also fast enough pulse rise time (see figure 1 and table 2). According to our estimations, in case a linear response better than 1% would be required up to 30 MeV energy, using the "LABRVD" voltage divider the PMT gain should be accordingly reduced and possibly halved with respect to the values used in this work.

Such results in terms of detector linearity are similar to those already reported in the literature [32, 37, 54, 55] where the PMTs have either smaller surface, or are underpowered or have an inherent lower nominal gain thanks to the reduced number of dynode stages.

As a last remark, it is important to point out that PMT behavior in terms of linearity is seldom reproducible (within 1% precision) so that actual detector energy response linearity might change, after PMT replacements [39].

**Figure 10**: *Left Panel: the measured linear response for LABRVD (filled point) and E1198-26 (empty squares) voltage dividers. Right Panel: the energy estimation error (residuals) for various high-energy γ rays following a linear detector calibration with low energy sources (see text). PMT gain was set in both cases so to obtain 30 mV pulse amplitude for a deposited energy of 661.6 keV (color online).*

## 4.4 Energy resolution

We estimated the energy resolution performance of large volume LaBr$_3$:Ce detectors with two different methods: i) a standard analog approach, based on shaping amplifiers and peak sensing ADCs and ii) a digital approach, based on free running ADC signal acquisition and subsequent digital processing.

The measurements with analog electronics were performed during the in-beam experiment at the ATOMKI Institute; we used an amplifier derived from the BaFPRO NIM module [49, 50] with bipolar shaping time of about 700 ns, followed by a peak sensing VME ADC (CAEN model V879) controlled by a specifically developed KMAX-based acquisition software [59, 60]. The energy of the measured γ rays ranged from 1 MeV up to 22.6 MeV (see table 3).

The measurements based on the digital approach were performed in the Milano Detectors Laboratory, a much more controlled environment inside the Physics Department of "Università degli Studi di Milano". Detector pulses were acquired using a 400 MHz, 5 GHz sampling frequency oscilloscope (LeCroy Waverunner 44X1). Because of very high energy dynamic range required and the poor ADC resolution (8 bits only), the analog front-end gain of the oscilloscope was adjusted from time to time to best fit the amplitude of the specific γ–ray pulses of interest (from 5 keV up to 9 MeV). The estimation of the released energy was performed using a straightforward box-car integration algorithm (over 250 ns) with the additional subtraction of the pulse baseline level (calculated over 250 ns).

We used two different 3.5" x 8" $LaBr_3$:Ce crystals during the tests (S/N K628CS_B at the ATOMKI and S/N M0249CS_B in Milano), coupled to R10233-100SEL PMTs equipped with "LABRVD" active voltage dividers.

According to Saint-Gobain Crystals' datasheets, the two crystals should provide almost equivalent energy resolution at 661 keV (3.0% FWHM for S/N K628CS_B operated at 658 V and 3.1% for S/N M0249CS_B operated at 696 V, measured with XP3540FLB2 PMTs and standard voltage dividers). Gain variance of 0.4% along the 8" axis of the crystal with S/N M0249CS_B was additionally quoted, while no corresponding information was provided for S/N K628CS_B crystal.

We estimated the FWHM energy resolutions of the two detectors as a function of γ–ray energy. Results are reported in figure 11 for the analog and in figure 12 for the digital approach. In both cases, the energy resolution of the $LaBr_3$:Ce detectors deviates from a strictly statistical behavior, i.e. $E^{1/2}$ asymptotic curve, in case of high-energy γ rays showing that the energy resolution of $LaBr_3$:Ce detectors tends to saturate at constant value around 0.5-1%, as already reported in the literature [32, 37].

This limitation is more likely to appear when pair production is the major γ–ray interaction mechanism, namely when a large fraction of the crystal volume is likely to be involved in the absorption process. This limit in the relative precision of energy estimation can be modeled as a simple linear dependence of the energy resolution with respect to the measured γ–ray energy. We then tried to interpret our experimental results for energy resolution in figures 11 and 12 not just in terms of the two basic contributions, i.e. electronic noise and quantum generation noise in the scintillation crystal, but also taking into account a third contribution due to energy resolution saturation:

$$ER_{FWHM} = \sqrt{a + bE + cE^2}. \quad (1)$$

In the previous equation, the first term '$a$' represents the total amount of noise unrelated to the measured energy, namely the equivalent noise charge of measurement (electronic noise); the second term '$b$' modulates the contribution of the statistical generation noise, while the third term '$c$' accounts for all the previously mentioned gain drift effects. We fitted the two experimental datasets using Eq. 1 and obtained the results summarized in table 4.

**Table 4:** *The fitting values for the three parameters (a, b, c) in Eq. 1, for the dataset of figure 11 (obtained with analog electronics) and the dataset of figure 12 (obtained with digital approach).*

We additionally required that the "$b$" values, accounting for the generation noise contribution to the energy resolution, must be a priori equal for the two scintillators. The equivalent noise charge, represented by the "a" values, turns out to be much higher for the "analog" acquisition, as a result of a much more complex and partially not optimized acquisition set-up. From figures 11 and 12 the importance of the third term in Equation 1, for energies above 2 MeV, is evident, as experimental data do not follow anymore the statistical behavior.
It is still unclear whether such a behavior is mainly caused by crystal non-homogeneity (resulting in light yield fluctuation and hence in signal gain fluctuation) rather than being determined by small, unnoticed changes in high voltage power supply level or even PMT gain fluctuations due to temperature changes (as a matter of fact, high-energy γ–ray measurements usually last for several hours). In order to better understand the origin of the energy resolution limit, additional and systematic measurements should be performed in a controlled environment, with a higher number of detectors, state-of-the-art high voltage power supplies and processing electronics.

As a direct evidence of the inherent quality of the energy resolution of large volume 3.5" x 8" LaBr$_3$:Ce detectors, figure 13 shows two low energy γ-rays spectra obtained with the acquisition setup in Milano, using $^{137}$Cs, $^{152}$Eu and $^{133}$Ba sources. Specifically, the inset spectrum shows the 5.6 keV and the 37.4 keV X-ray peaks of $^{138}$Ba [38] and the corresponding FWHM energy resolutions of 1.8 keV and 5.3 keV respectively, thus proving the excellent detector performance even for such low energy X rays. Figure 14 shows the energy spectrum measured at the ATOMKI Institute in Debrecen with the S/N K628CS_B LaBr$_3$:Ce detector, the "LABRVD" active voltage divider and the analog electronics, in case of 17.6 and 22.6 MeV monochromatic γ rays (see table 3). It is worth noticing the much higher detector efficiency (as compared, for example, with that of 2" x 2" LaBr$_3$:Ce detectors [32]), the complete separation between the full energy and the first escape peak and the complete absence of the second escape peak. As already mentioned, LaBr$_3$:Ce is nowadays the only scintillator able to separate the full energy peak from the first escape peak, up to at least 25 MeV γ–ray energy.

**Figure 11**: *The FWHM energy resolution in the energy range between 1 and 22.6 MeV obtained at the ATOMKI Institute, with the S/N K628CS_B LaBr$_3$:Ce detector coupled to the "LABRVD" active voltage divider and the analog electronics. The continuous line represents the complete function of Eq. 1 with the corresponding fitting parameters of table 4, while the dashed line represents only the first two contributions associated to the 'a' and 'b' parameters, namely the electronic noise and the statistical contribution.*

**Figure 12:** *The FWHM energy resolution in the energy range between 5 keV and 9 MeV measured with the S/N M0249CS_B LaBr$_3$:Ce detector coupled to the "LABRVD" active voltage divider, free running ADCs and digital processing. The continuous line represents the complete function of Eq. 1 with the corresponding fitting parameters of table 4 while the dashed line represents only the first two contributions associated to the 'a' and 'b' parameters, namely the electronic noise and the statistical contribution.*

**Figure 13**: *The two low energy spectra obtained with the digital acquisition setup, using $^{137}$Cs, $^{152}$Eu and $^{133}$Ba sources, measured with the 3.5" x 8" LaBr$_3$:Ce detector and the active voltage divider. The leftmost energy peak (at 37.4 keV) corresponds to the X-ray K shell of Ba, while the 81 keV energy peak comes from the $^{133}$Ba source. In the inset spectrum it is also shown the 5.6 keV peak corresponding to the X-ray L shell of Ba.*

The top panel of figure 15 shows the energy spectrum measured with the S/N M0249CS_B LaBr$_3$:Ce detector, the "LABRVD" active voltage divider, free running ADCs and digital processing, using the Am-Be-Ni source, up to 9 MeV γ-ray energy (see section 4.1). As an additional comparison, the bottom panel of figure 15 shows the same energy spectrum acquired using a HPGe detector with 80% detection efficiency with respect to a 3" x 3" NaI scintillator.

**Figure 14**: *The energy spectrum measured at the ATOMKI Institute, with the S/N K628CS_B LaBr$_3$:Ce detector, the "LABRVD" active voltage divider and the analog electronics in case of 22.6 MeV (top panel) and 17.6 MeV (bottom panel) monochromatic γ rays (see table 3).*

**Figure 15**: *Top panel: the energy spectrum measured with the S/N M0249CS_B LaBr$_3$:Ce detector, the "LABRVD" active voltage divider, free running ADCs and digital processing, using the Am-Be-Ni source (see section 4.1). Bottom panel: as a reference, the same energy spectrum measured using a HPGe detector.*

## 4.5 Time resolution

The time resolution obtainable with LaBr$_3$:Ce detectors is not uniquely determined by the crystal properties. Sometimes, more technology-related aspects may eventually act as limiting factors, e.g. the rise-time of the detector signals, the choice of the photo-detector (PMT, APD, etc), of the voltage divider (in case of PMTs) and of the applied high voltage power supply level (see figure 1) and, finally, the quality of the time pick-off electronics.

We noticed, for example, that in case of LaBr$_3$:Ce detector signals, the time resolution obtainable with constant fraction discrimination (CFD) modules significantly changes not only with the time delay implemented, as discussed in [8], but also among the various models provided by different manufacturers.

The time resolution measurements were performed with an ORTEC CF8000 CFD, using a $^{60}$Co source and setting the CFD lower threshold limit at about 1 MeV. Slightly worse, but still comparable results were obtained with a digital CFD and the BAFPRO module of ref [49,50].

Coincidence-based measurements were performed using an additional 3" x 2" (76mm x 50mm) hexagonally shaped HELENA [61] BaF$_2$ scintillator as the reference time detector, providing 370 ps FWHM intrinsic time resolution (with 2ns time delay), while the electronics provided better than 50 ps FWHM intrinsic time resolution.

We tested the four LaBr$_3$;Ce detector configurations, reported in table 5, with the associated PMT, voltage divider, high voltage power supply level and CFD time delay. It is worth mentioning that cases 1, 2 crystals (see table 5), unlike the other two, came within a sealed capsule and that the two CFD time delay values of case 3, 4 (see table 5) were not likely to be the optimum ones but rather the maximum ones allowed by the CFD module.

**Table 5:** *The four detector configurations that we tested for time resolution, with the associated crystal size, PMT, voltage divider and CFD time delay. The last column lists the measured time resolution. The estimated error in the measured FWHM is 35 ps.*

The intrinsic time resolutions measured for the four detector configurations are reported in table 5. As expected from the rise time measurements of section 2, as a general trend the intrinsic time resolution worsens with increasing crystal volume, as already observed in the literature for scintillation crystals up to 2" x 3" in size [1,7]. Indeed, two effects are present: first, the rise-time of the average detector pulse increases, because of more light reflections inside the crystal and slower PMT rise-time, as reported in section 2 (see figures 1 and 2), secondly the individual pulse lineshapes are subject to fluctuations because of the random light reflections.

However, we were still able to obtain an intrinsic time resolution slightly better than 1 ns FWHM for our 3.5" x 8" detectors, which is of course a worse result with respect to the intrinsic time resolution of LaBr$_3$:Ce, obtainable for example with a 1"x1" crystal (230 ps [14]), but nonetheless still acceptable for most applications.

## 5. Simulated and measured efficiency

The absolute γ–ray detection full energy peak efficiency of 3.5" x 8" large volume detectors was estimated by means of the 'sum peak' technique [58]. We used a $^{60}$Co source positioned at a distance of 10 mm (± 1 mm) from the detector front face (see figure 16). In case of $^{60}$Co, the 'sum peak' technique is based on the comparison of the energy spectrum counts in the two full energy peaks at 1173 keV and 1332 keV, against the counts in the 'sum peak' line (at 2505 keV). This method relies on the assumption that the two detector efficiencies (at 1173 keV and 1332 keV) have almost equal value, which is a very reasonable assumption in case of very large volume detectors. Figure 16 shows the experimental results, together with the GEANT3 simulation results (from 100 keV up to 30 MeV), very well in agreement with each other.

We can then use the predictive power of the simulations to reproduce the efficiency for a source positioned at 200 mm from the detector front-face, as shown in figure 17. Note that absolute detection efficiency at arbitrary distance from the detector front-face is not easily determinable by means of simple scaling factors, as it depends on the solid angle subtended by the detector, the γ–ray entrance angle and the γ–ray energy, namely on the main mechanism of γ–ray interaction.

**Figure 16**: *Simulated and experimentally measured values of absolute full energy peak efficiency for a large volume 3.5" x 8" LaBr$_3$:Ce detector, with a $^{60}$Co source positioned 10 mm away from the detector surface. The inset plot shows a magnified portion of the main graph up to 5 MeV energy range.*

**Figure 17**: *Simulated values of absolute efficiency for a large volume 3.5" x 8" LaBr$_3$:Ce detector, with a $^{60}$Co source positioned 200 mm away from the detector surface.*

## 6. Conclusions

In this work we reported the results of a series of tests performed on 3.5" x 8" LaBr$_3$:Ce scintillation detectors, among the largest ones available at the time of writing, evaluating many properties not directly scalable from the ones of smaller detectors.
Large volume LaBr$_3$:Ce scintillators are very promising detectors, to be used in combination, or in some cases even as an alternative to HPGe detectors. They may indeed provide very good results in case of high-energy γ–ray measurements, e.g. coming from the decay of highly collective nuclear states, radioactive beam facilities or both elastic and inelastic reactions in present and future radioactive and Nuclear Resonance Fluorescence (NRF) facilities like, for example, ELI-NP or HiγS [21-24]. The demonstrated capability to efficiently measure and separate the full energy peak from the first escape one for γ-rays up to at least 25 MeV is unique of large volume LaBr$_3$:Ce detectors.
We tested the LaBr$_3$:Ce detectors using two different voltage dividers coupled to the PMTs, a commercially available reference passive voltage divider (E1198-26) and the "LABRVD" active voltage divider, specifically designed in Milano . We showed that the use of the "LABRVD" allows the measurement of low-medium and high-energy γ rays (up to 18 MeV) with no need of reducing the number of PMT dynodes nor lowering the operating Voltage to values where the PMT performances could be compromised.
We compared the results obtained with the large volume crystals against the ones obtained with smaller detectors, either available from the literature or directly acquired in Milano, showing how the pulse shape changes with the crystal dimension, with consequent worsening of the intrinsic time resolution as the dimension increases, still remaining better than 1 ns FWHM for our 3.5"x8" detectors. We also evaluated the stability of performance as a function of the event count rate (< 5 keV @ 898 keV between 5 – 250 kHz), having disentangled the direct effect of the count rate itself from the secondary slower effect due to temperature change in the electronics due to count rate changes. As expected the active voltage divider can stand large variations of count rate without a significant change in gain and resolution making the detector more suitable for in-beam measurements.  We showed that the linearity of the detector response as a function of the interacting energy can be better than 1% for $E_\gamma$ up to 20 MeV and we have estimated that, by lowering the voltage, good linearity can be achieved also for γ-rays of much higher energy.
The Energy resolution limitation between 0.5% and 1% in case of high-energy γ rays, already observed in previous works, was confirmed. We were able to correct the energy resolution deviation from the statistical behavior at energies above pair production by introducing a term which takes into account this energy resolution saturation. We proved that LaBr$_3$:Ce detectors are anyhow able to clearly separate the full energy peak from first escape peak up to at least 25 MeV γ–ray energy, which is an unique feature for a scintillator detector up to now. Good detection efficiency were eventually verified, also with respect to dedicated GEANT3 simulations.

As the overall conclusion, large volume 3.5" x 8" LaBr$_3$:Ce scintillators are among the most promising detectors for basic research in nuclear structure.

**ACKNOWLEDGMENTS**

This work has been supported by the Hungarian OTKA Foundation No. K 106035. The work is supported by the TA´MOP-4.2.2/B-10/1-2010-0024 project. The project is co-financed by the European Union and the European Social Fund. This work was also supported by *NuPNET - ERA-NET* within the the NuPNET GANAS project, under grant agreement n° 202914 and from the European Union, within the "7$^{th}$ Framework Program" FP7/2007-2013, under grant agreement n° 262010 – ENSAR-INDESYS. The authors would like to thank G.De Angelis from Laboratori Nazionali di Legnaro for the 1"x1" + XP20D0B detector.

Figure 1

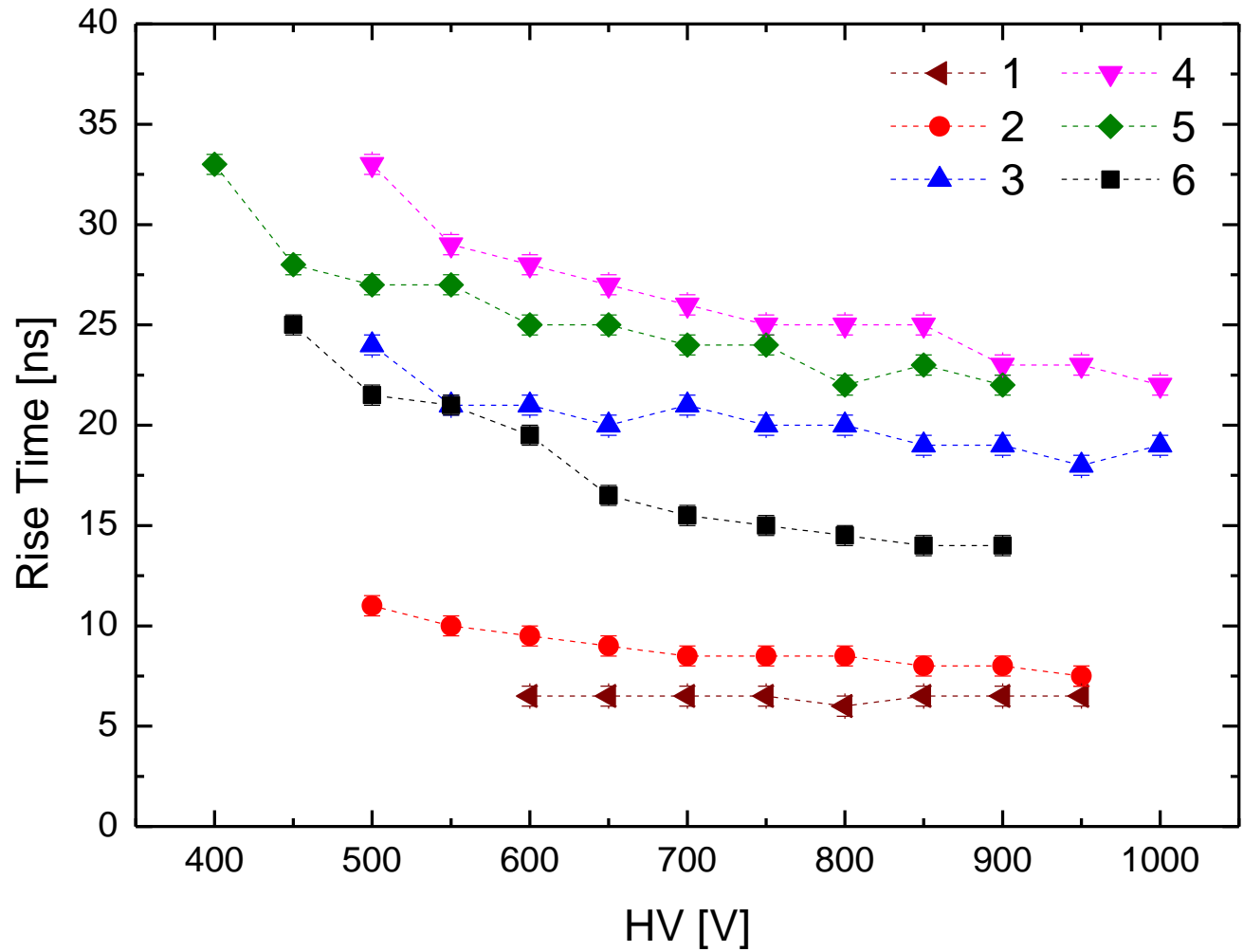

Figure 2

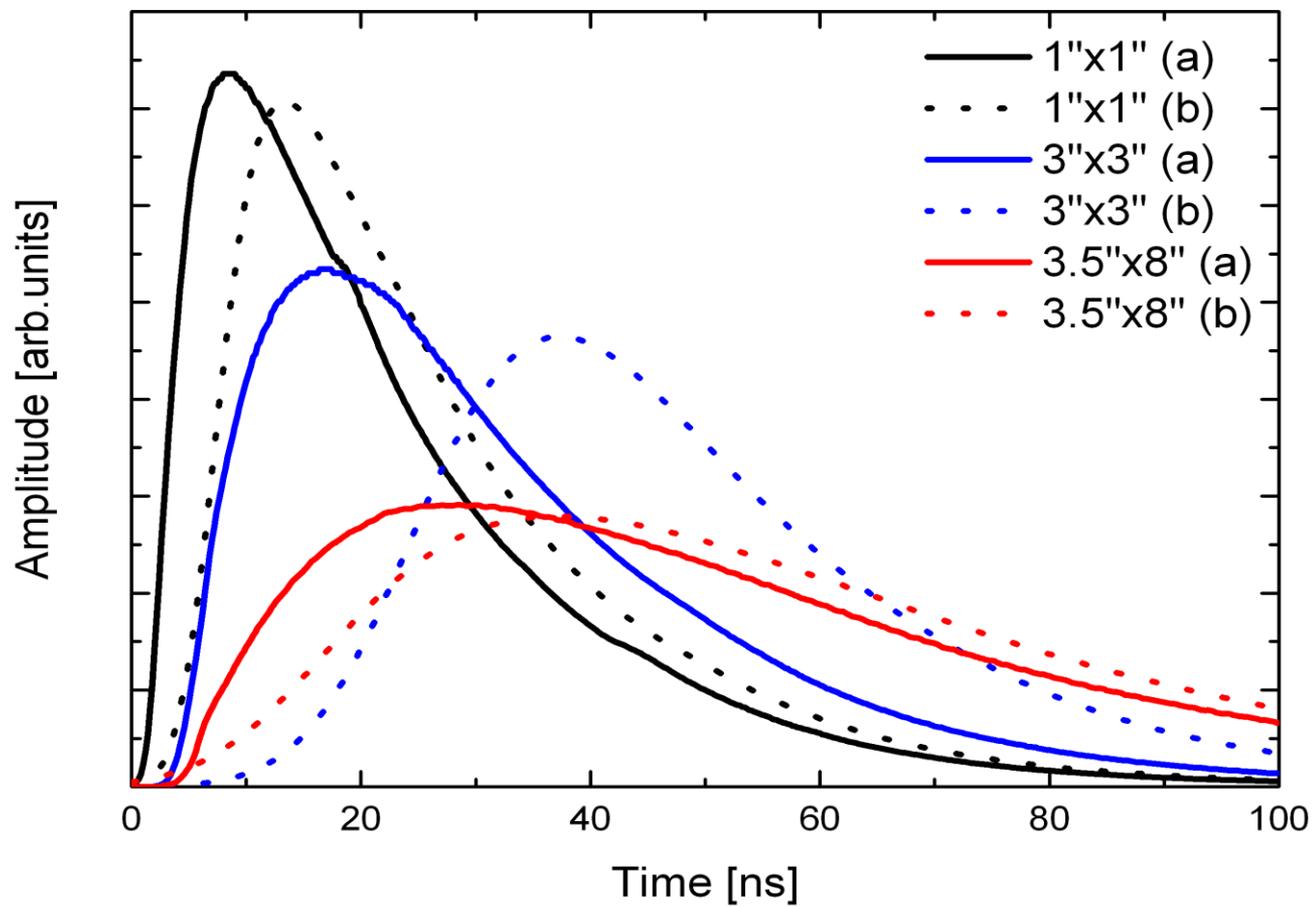

Figure 3

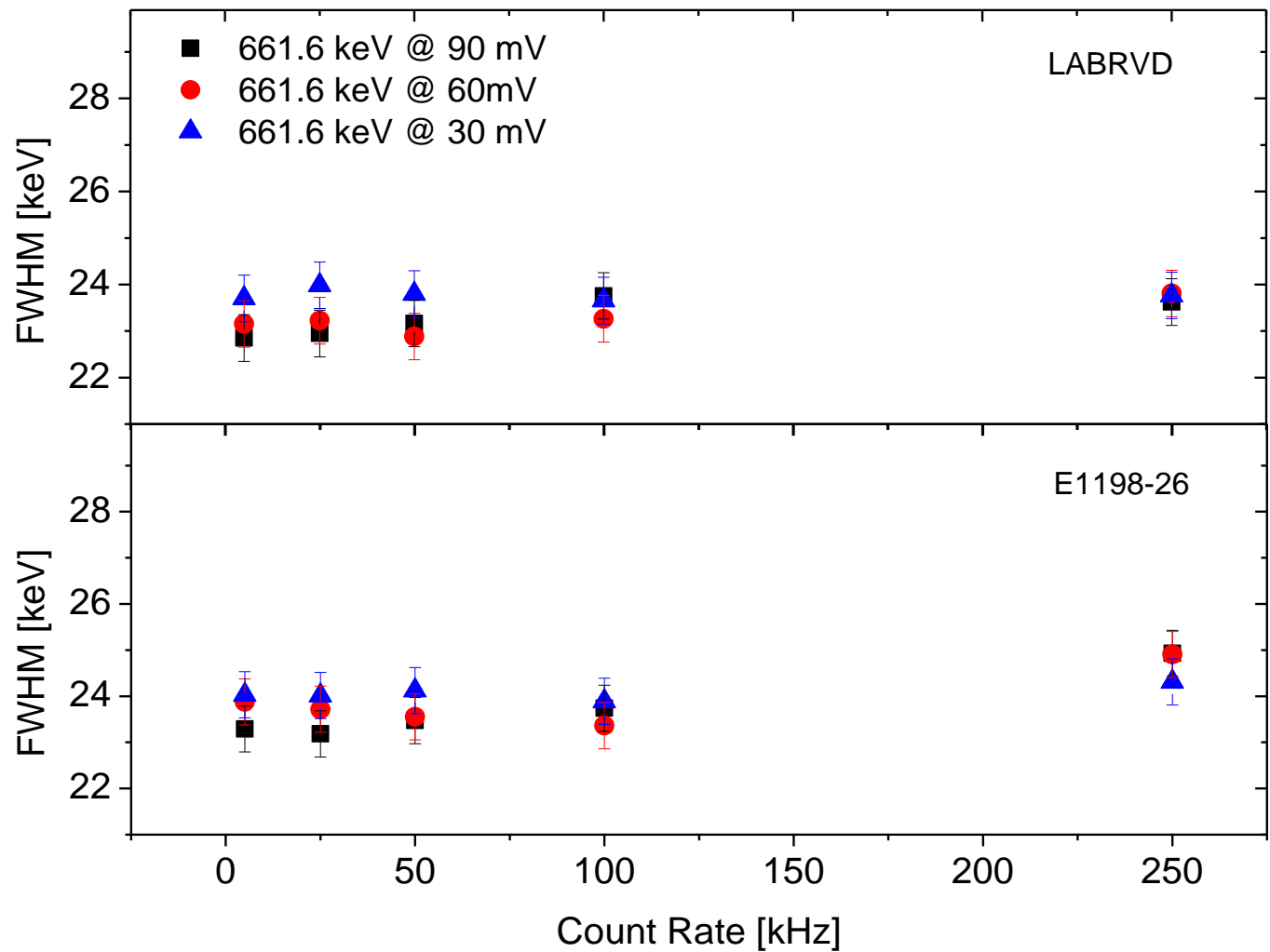

Figure 4

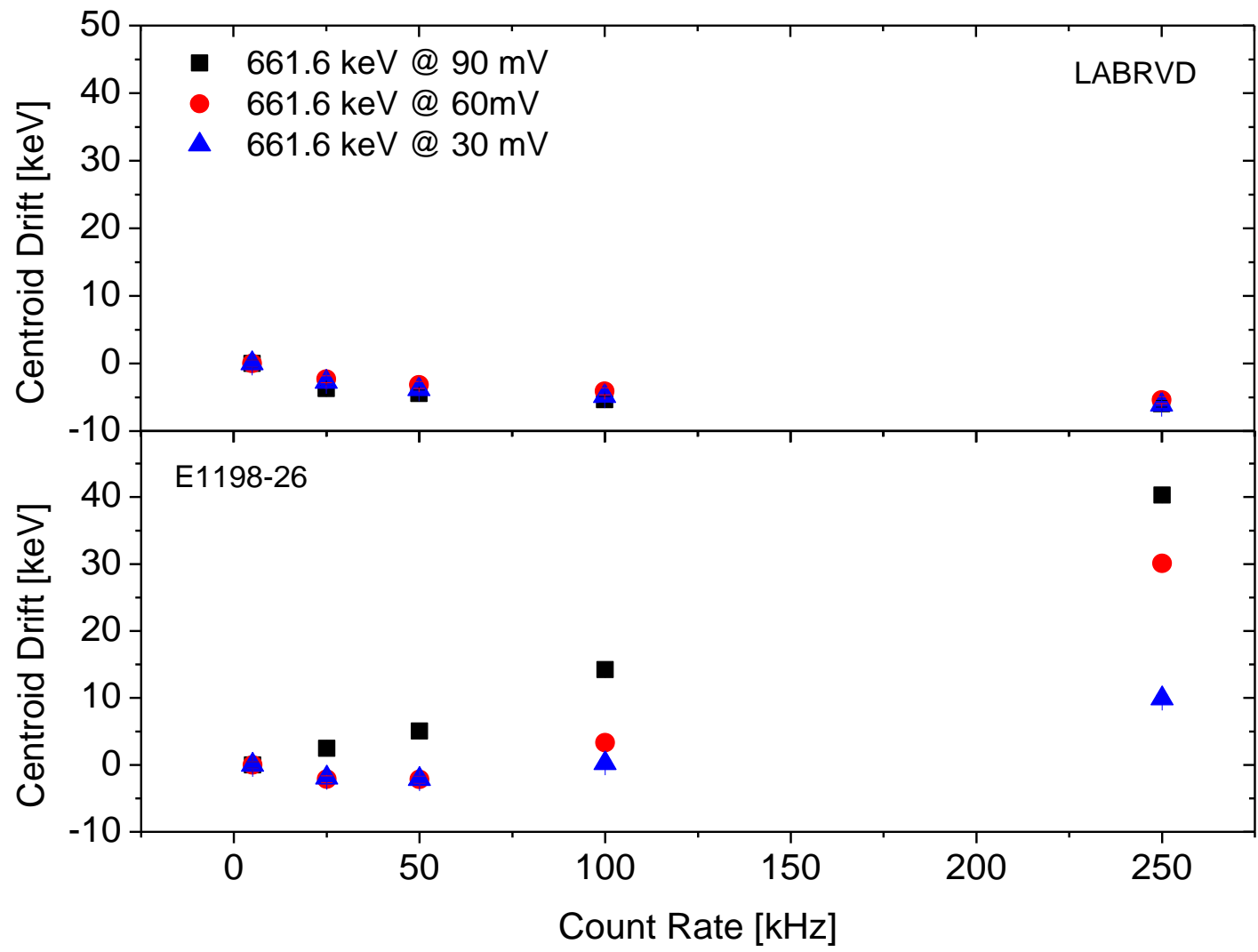

Figure 5

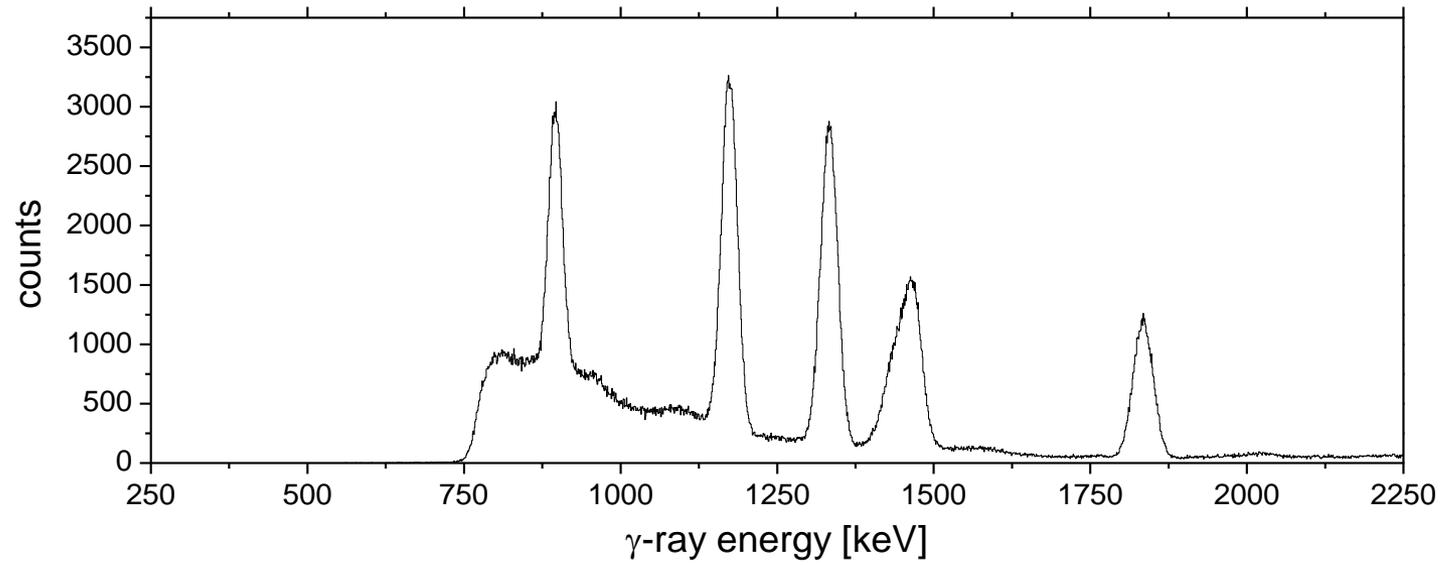

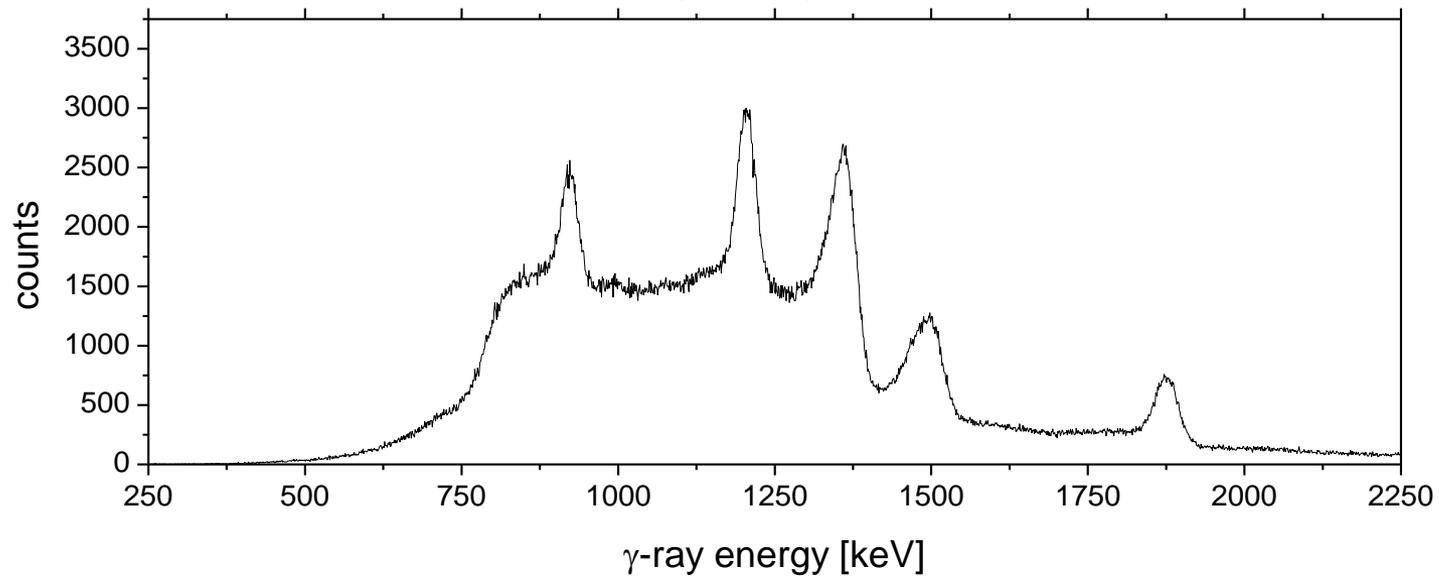

Figure 6

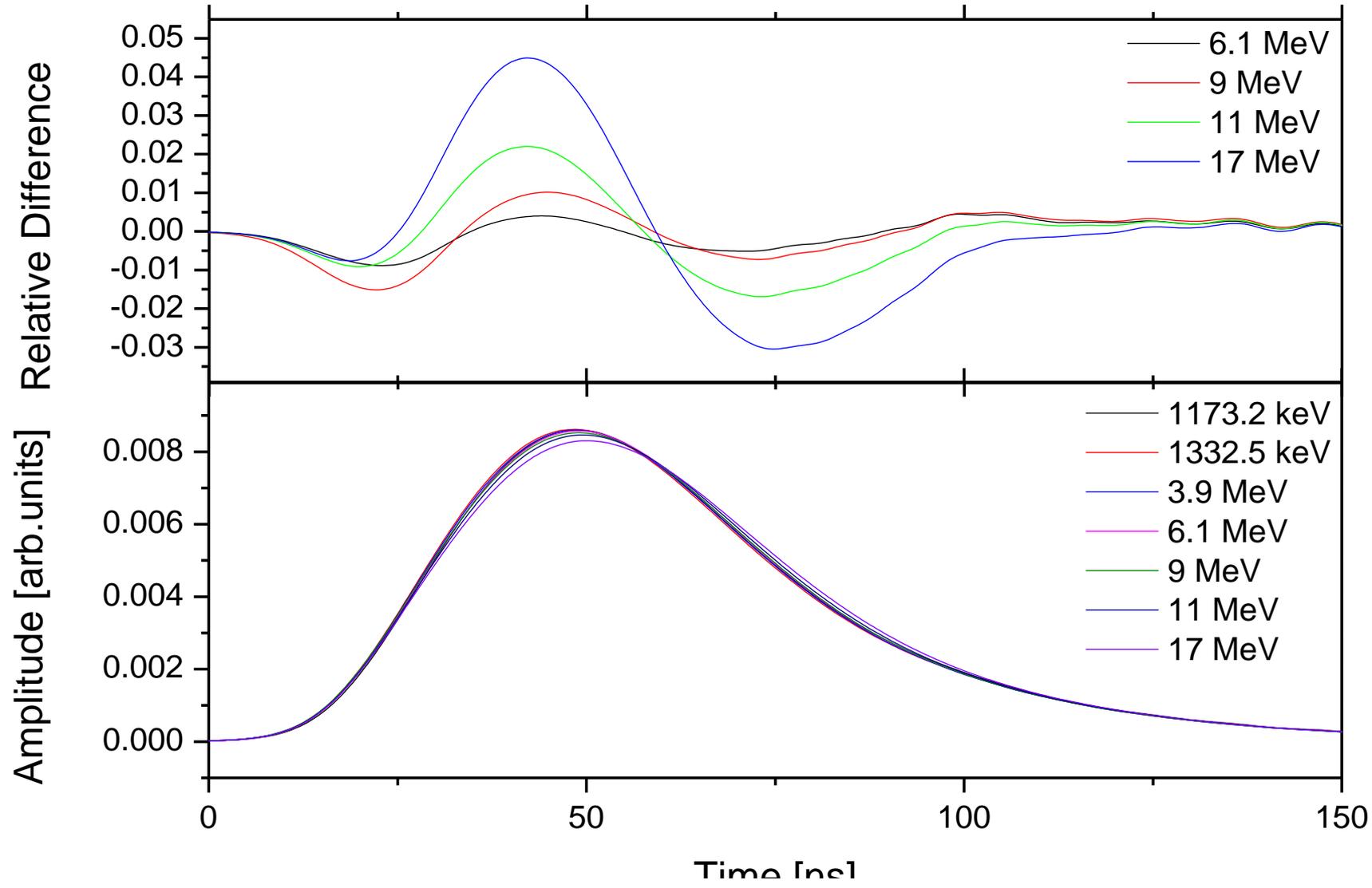

Figure 7

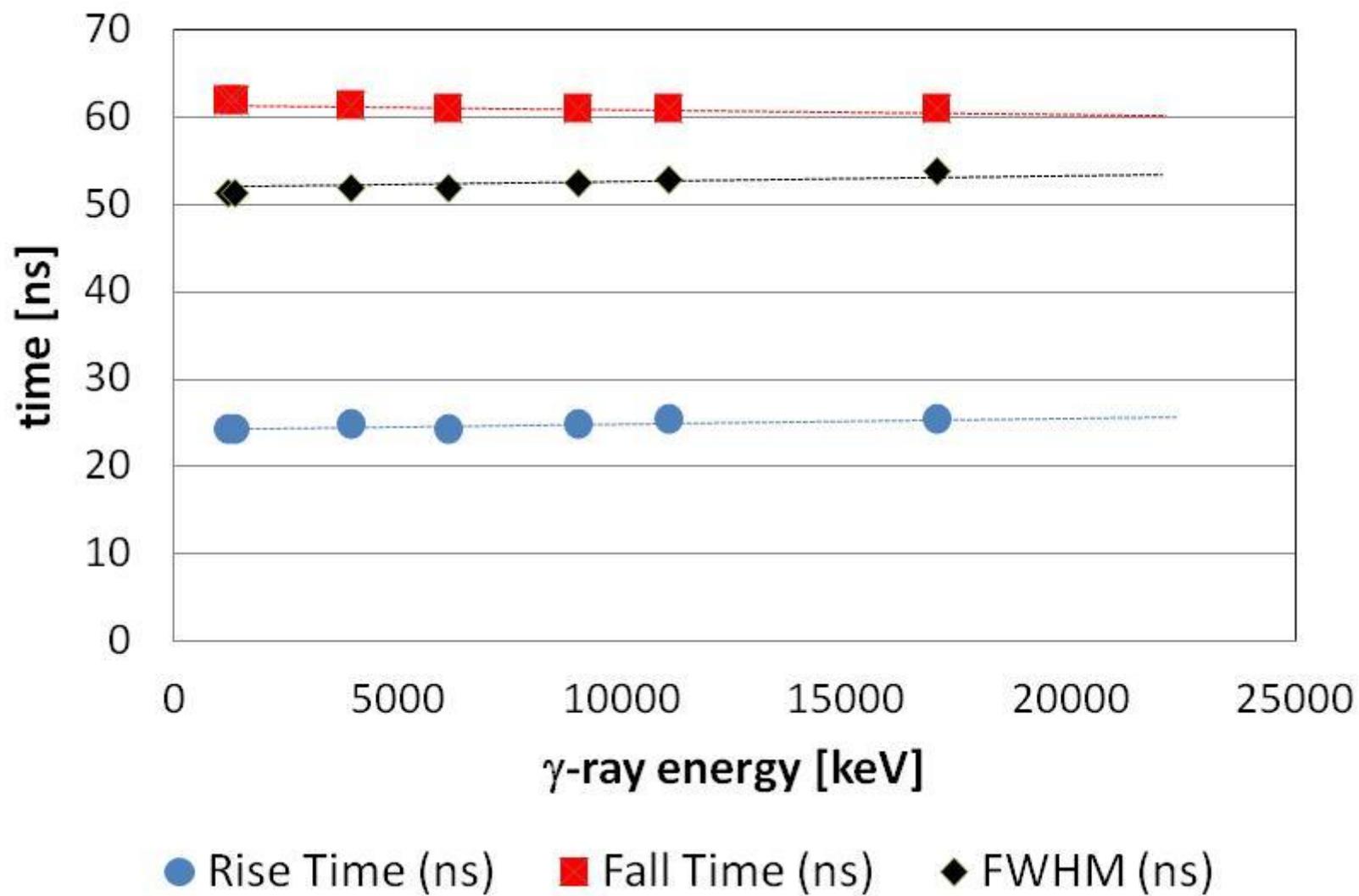

Figure 8

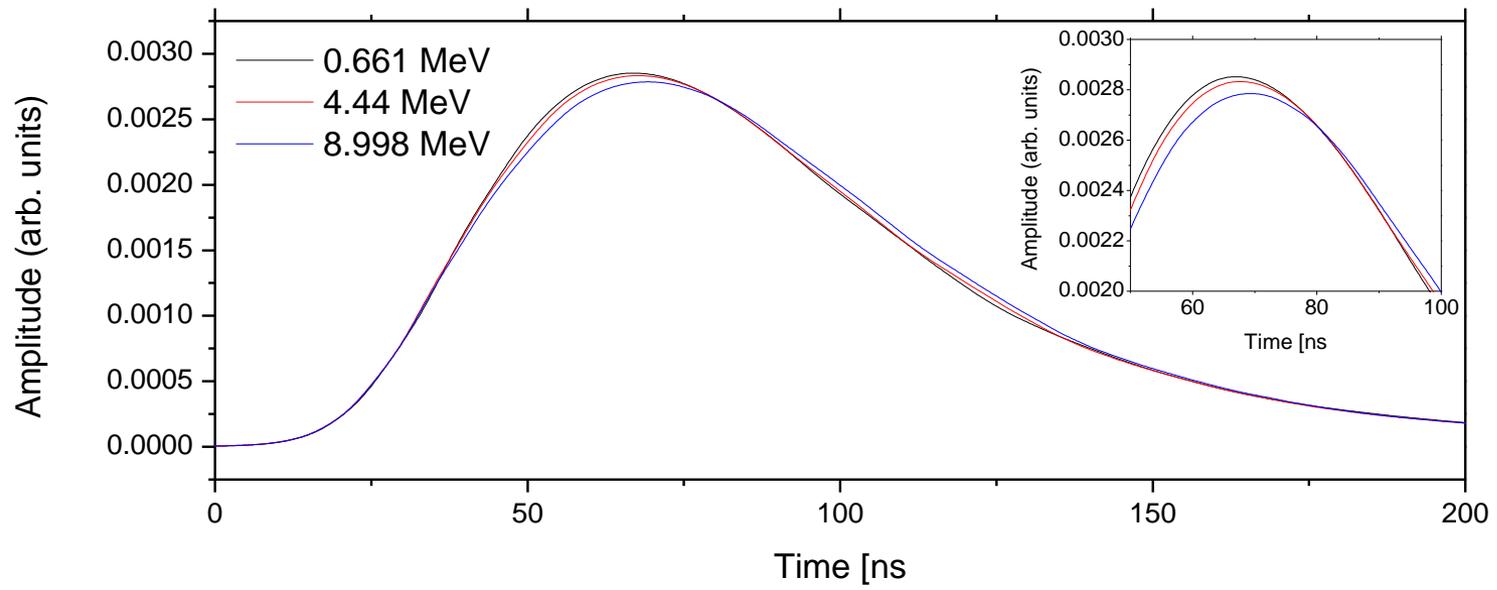

Figure 9

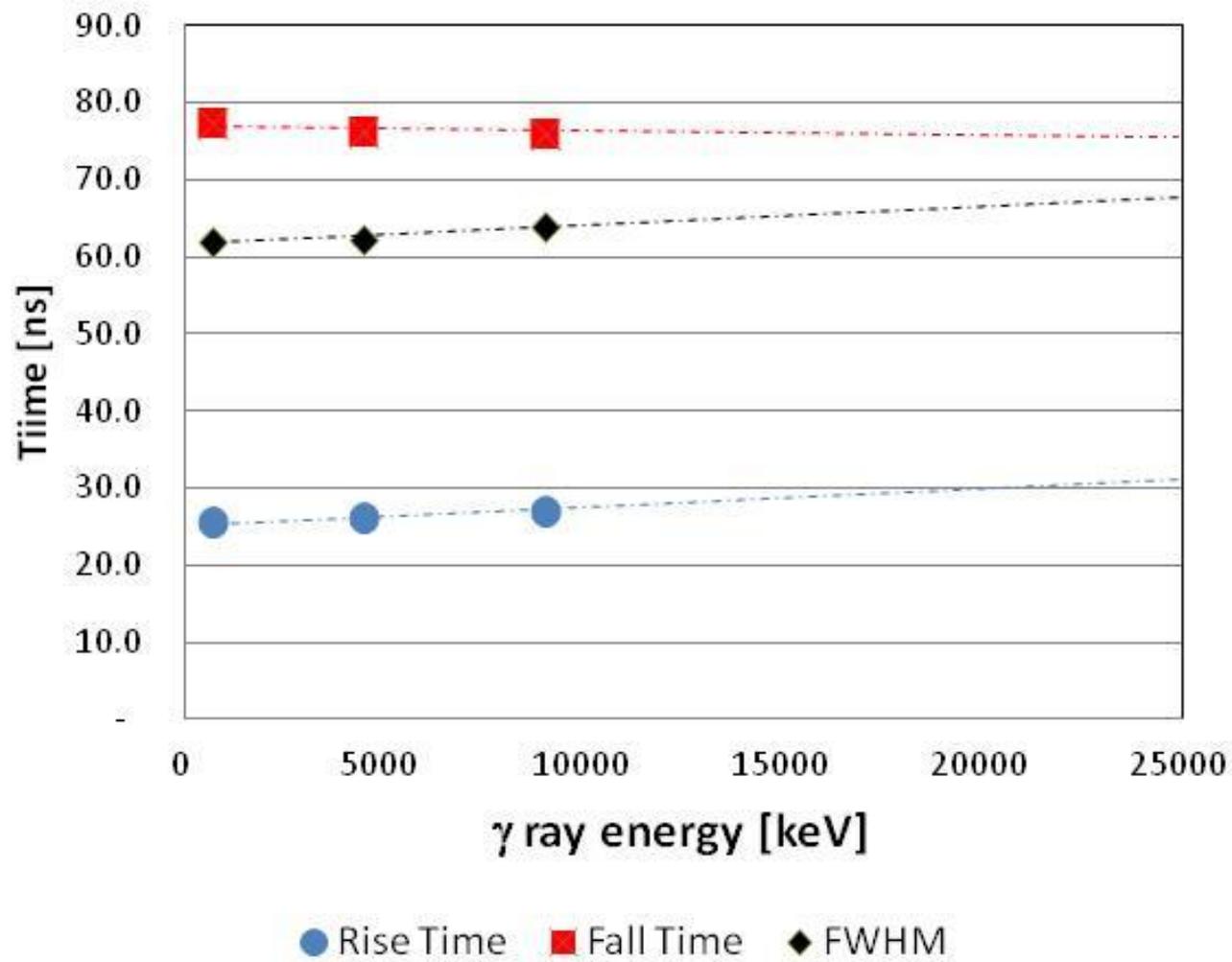

Figure 10

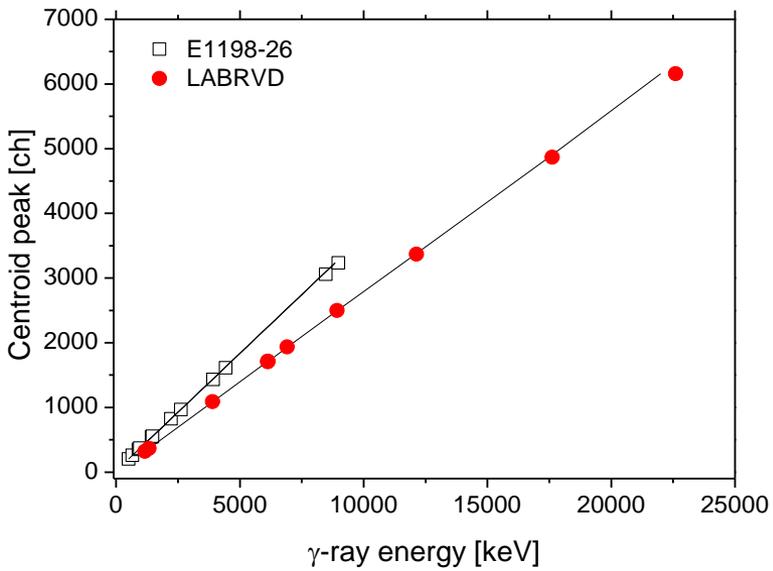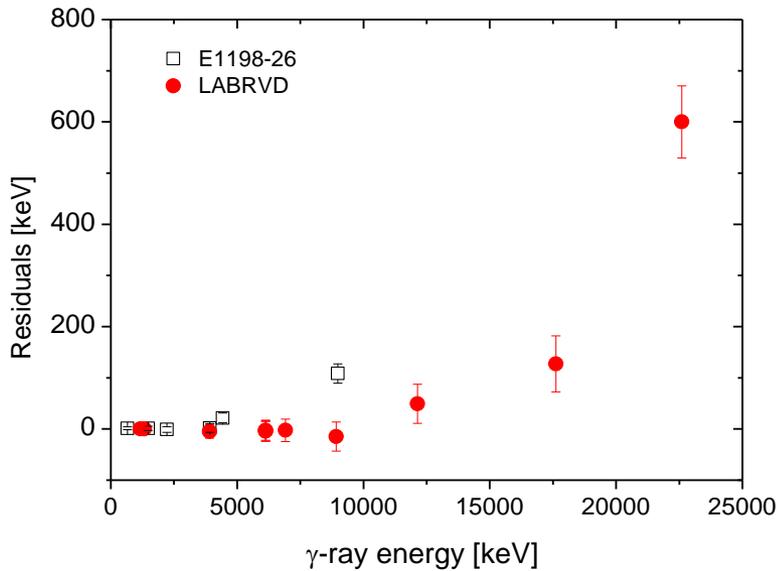

Figure 11

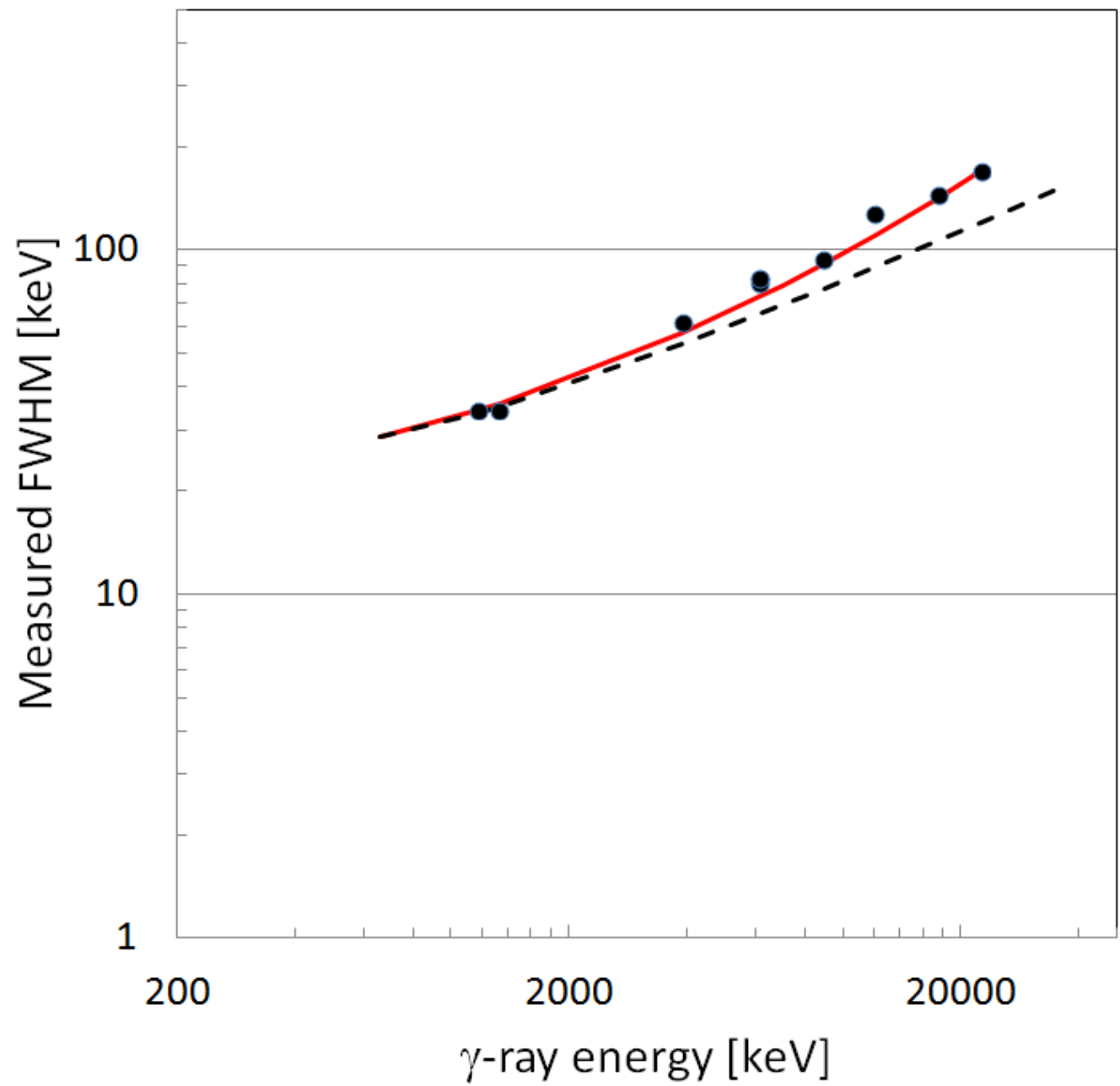

Figure 12

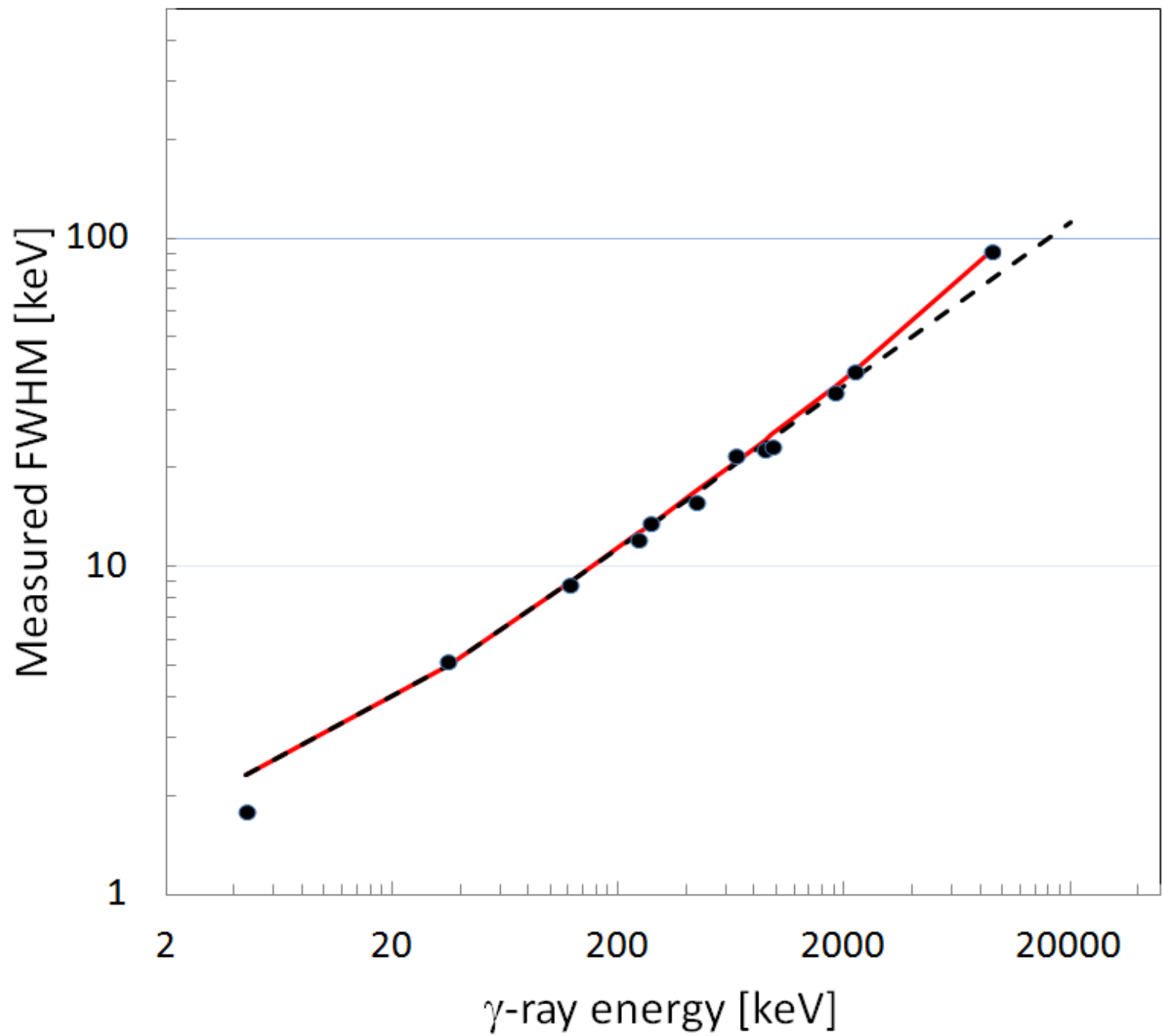

Figure 13

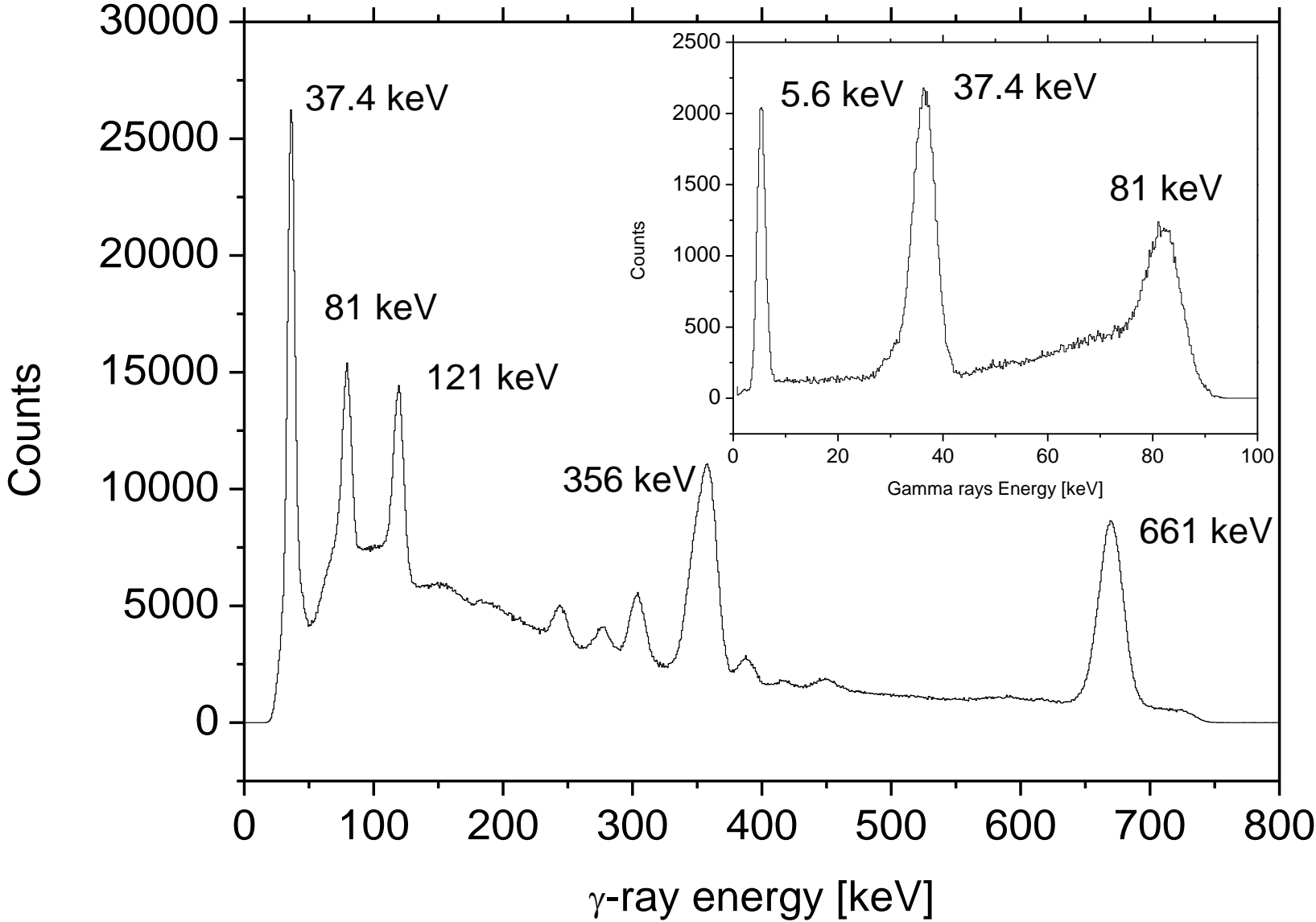

Figure 14a

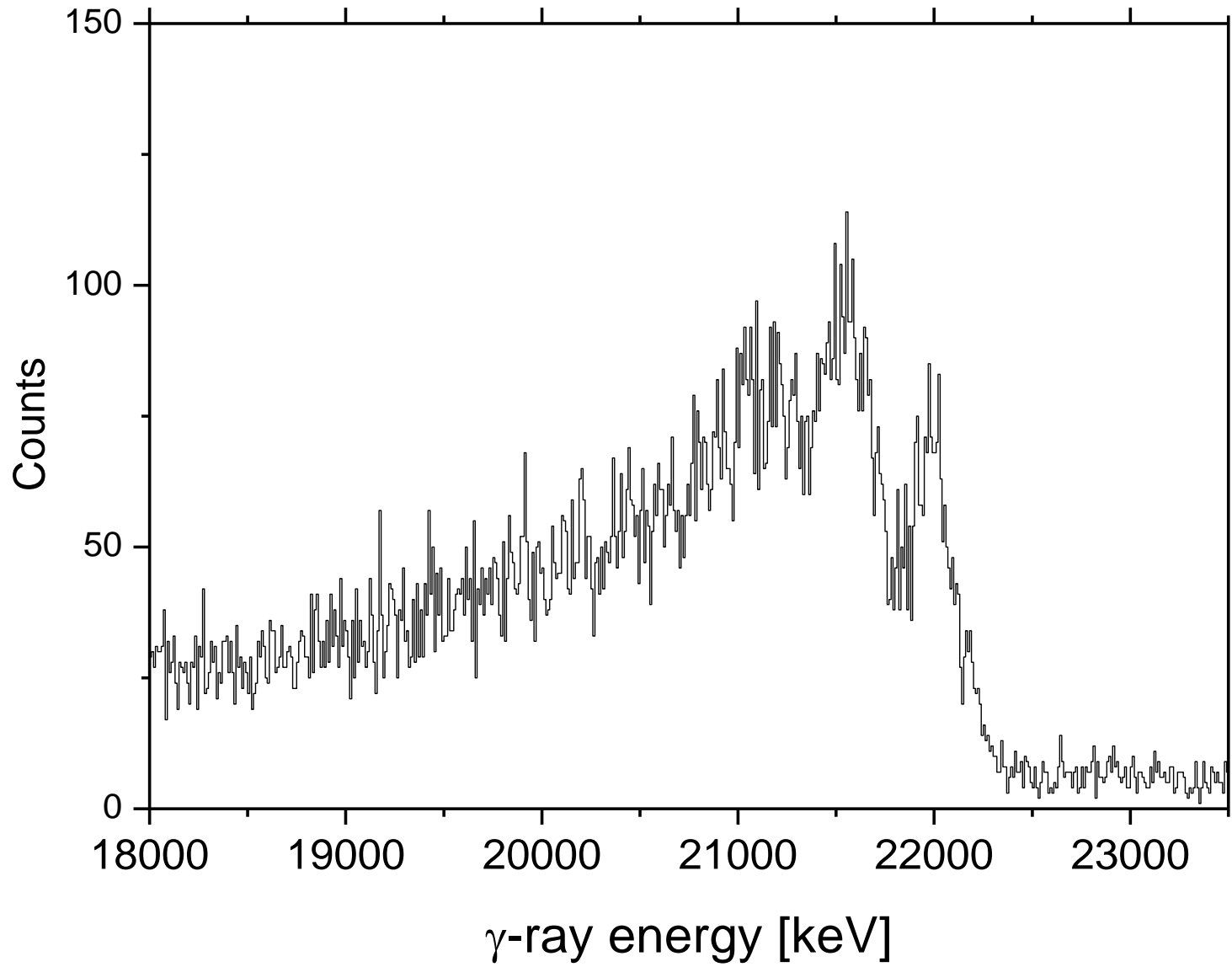

Figure 14b

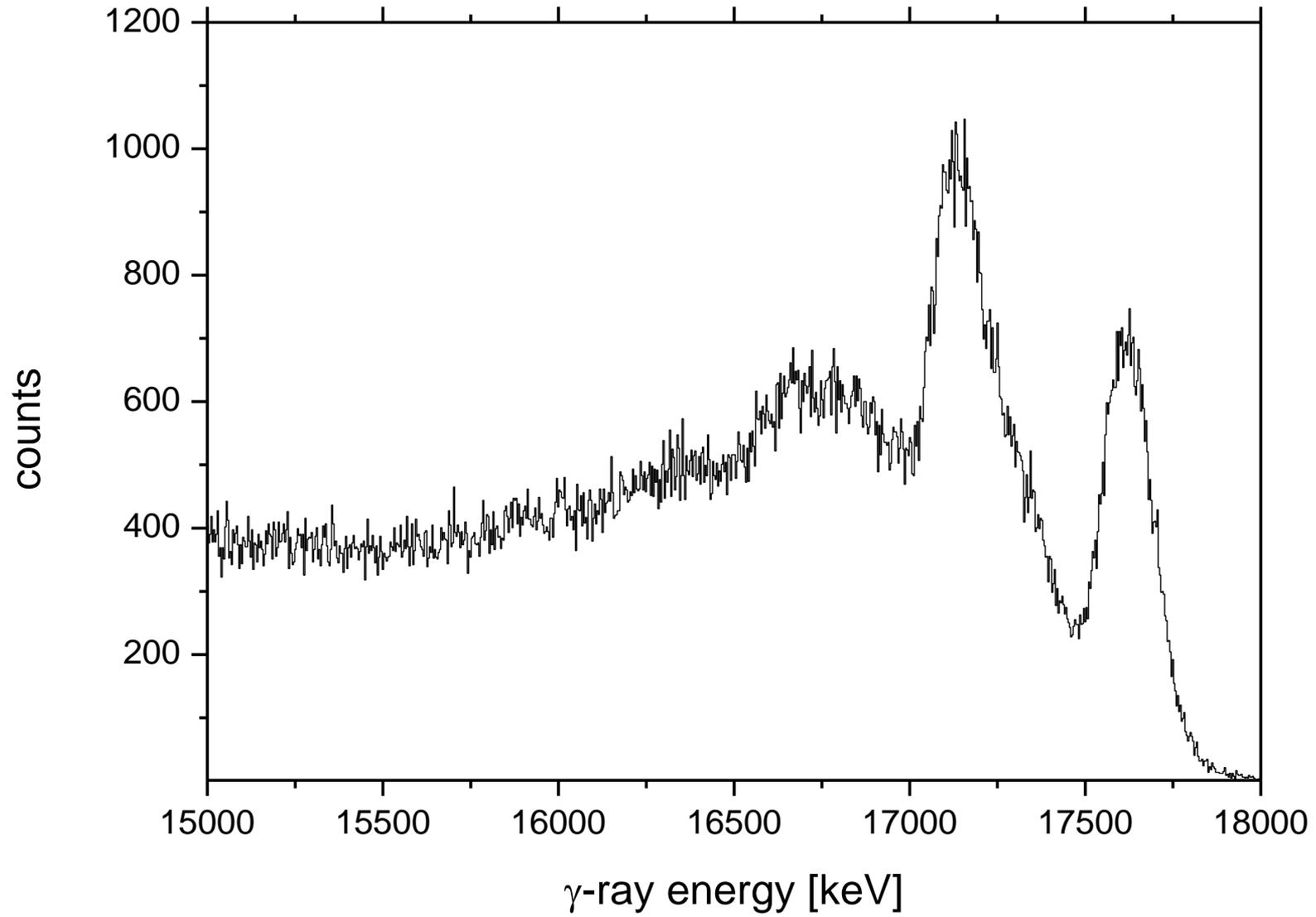

Figure 15

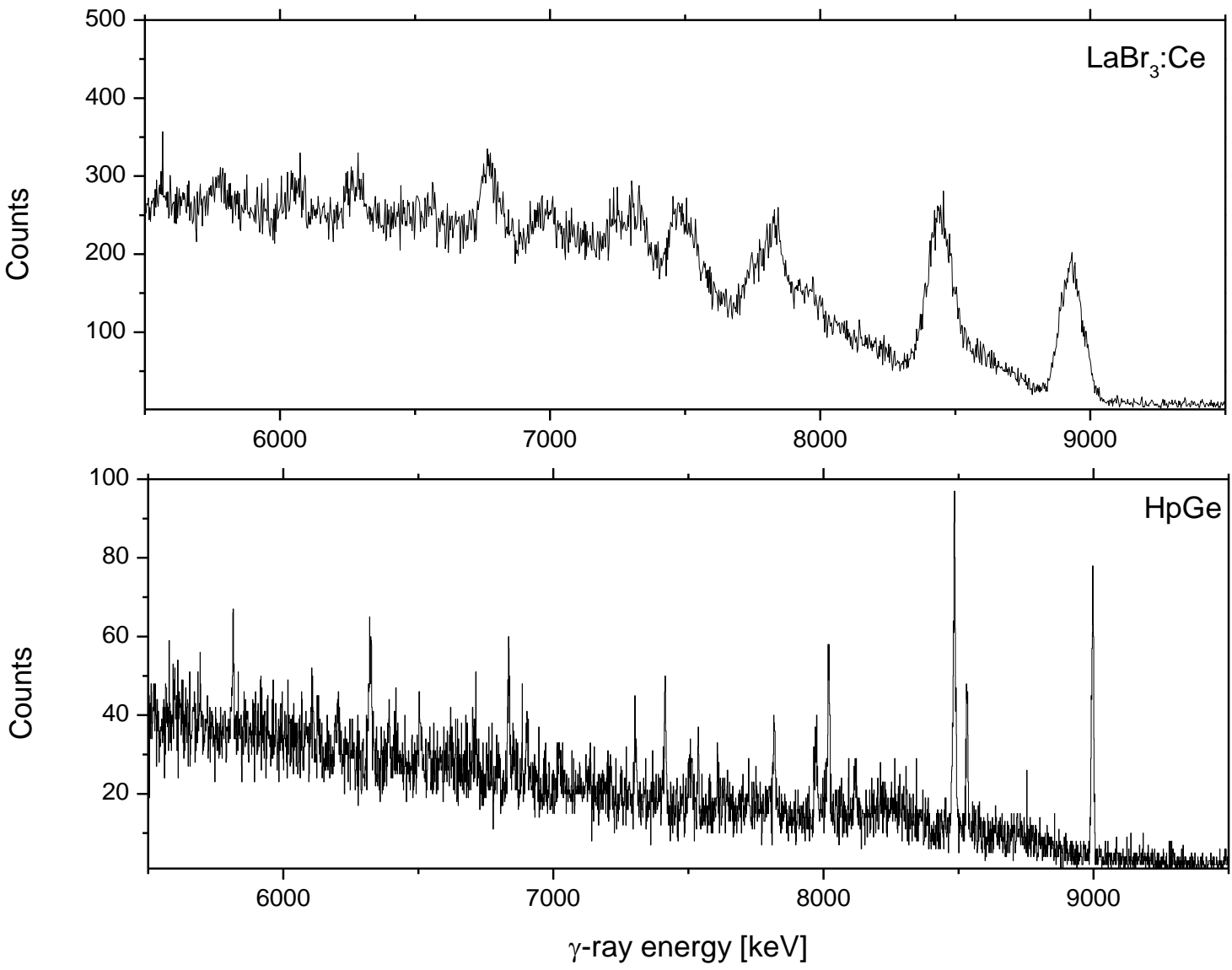

Figure 16

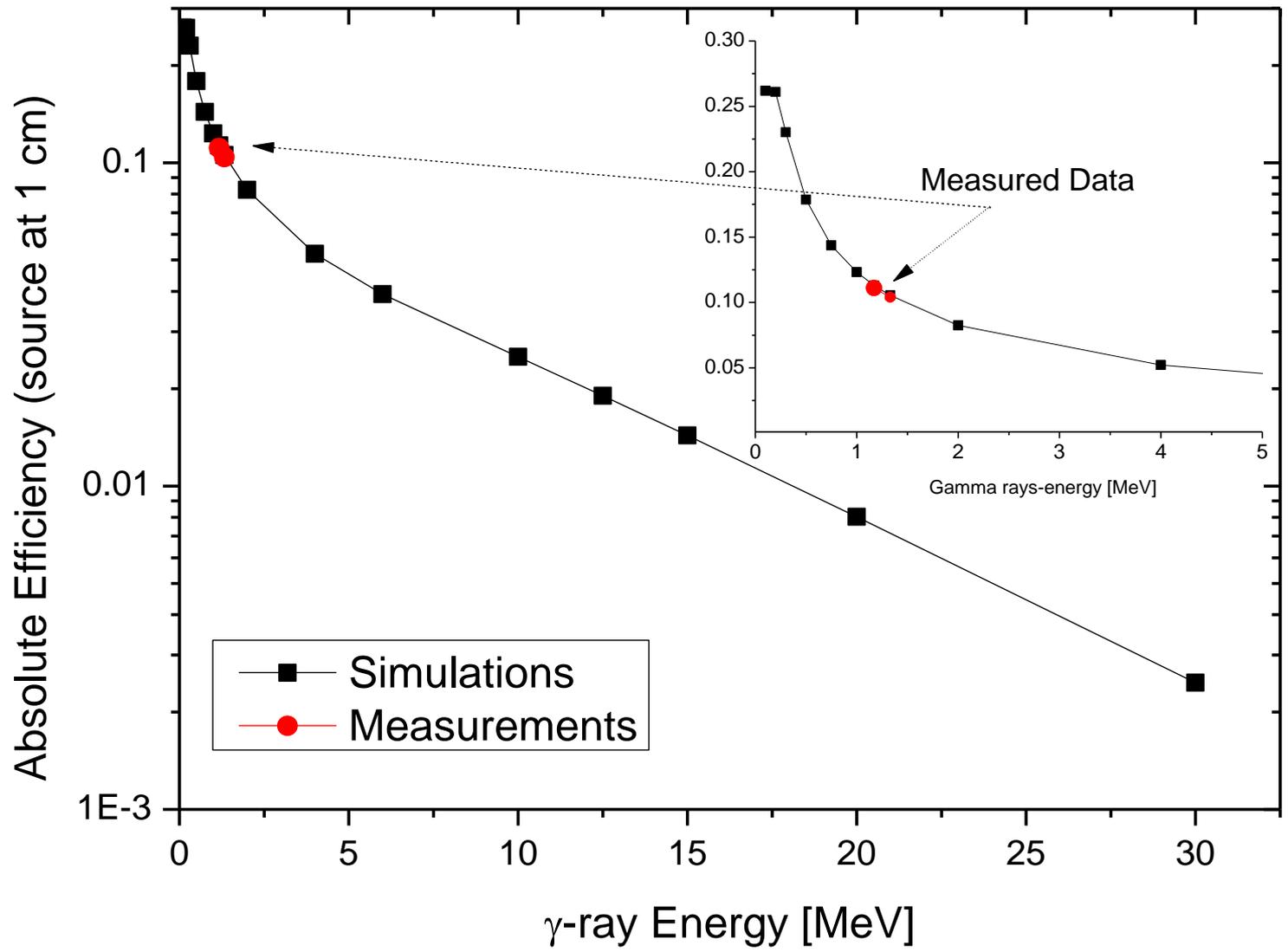

Figure 17

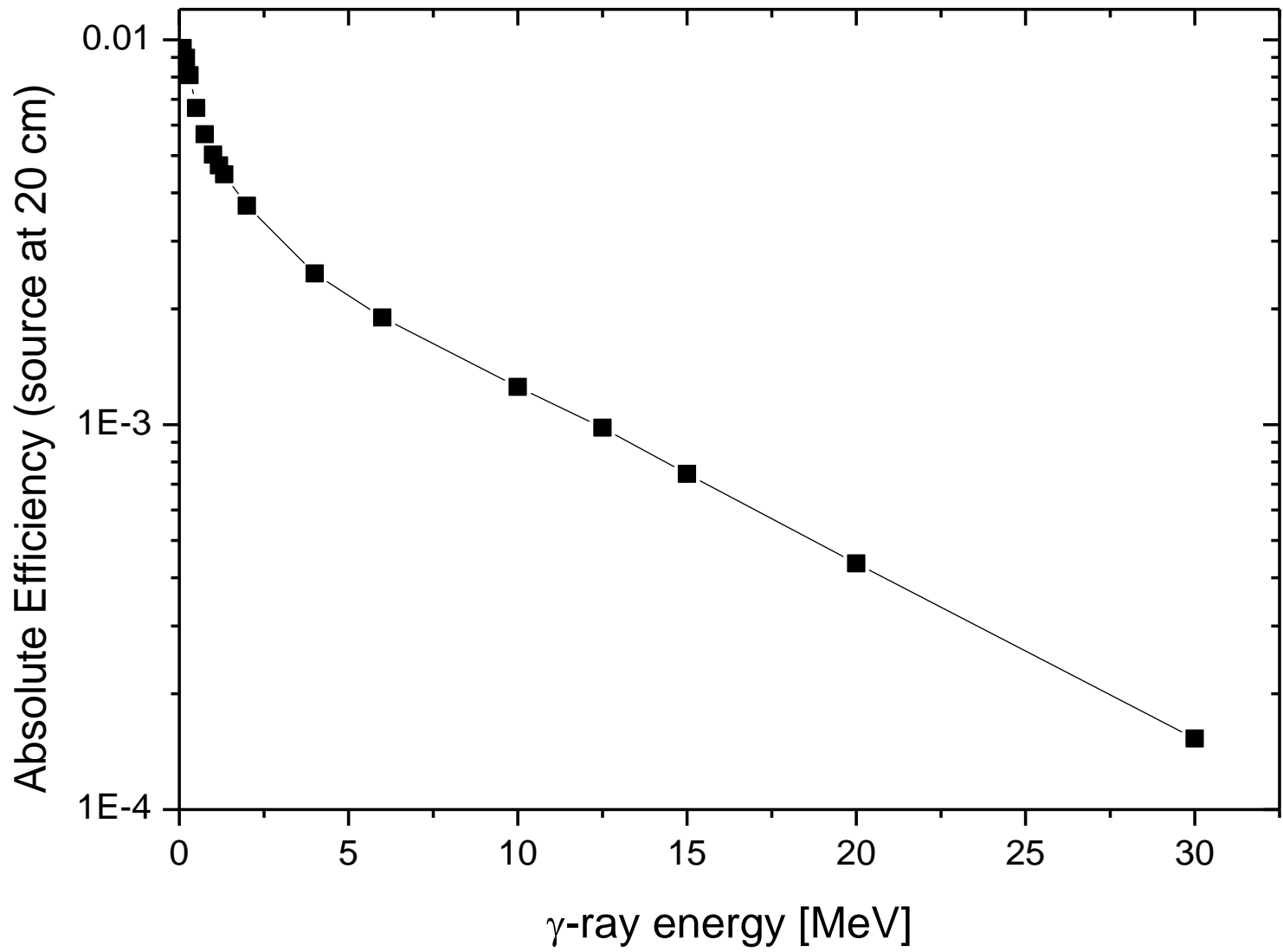

| # | Detector size | Associated PMT | Voltage Divider | Rise Time |
|---|---|---|---|---|
| 1 | 1" x 1" | XP 20 D0B | 184K/T * | < 3 ns |
| 2 | 1" x 1" | XP2060B | AS20 ** | 3 ns |
| 3 | 3" x 3" | R6233-100SEL | LABRVD * | 6 ns |
| 4 | 3.5" x 8" | R10233-100SEL | LABRVD * | 10 ns |
| 5 | 3.5" x 8" | R10233-100SEL | E1198-26 * | 10 ns |
| 6 | 1" x 1" | R10233-100SEL | LABRVD * | 10 ns |

*Negative Voltage
**Positive Voltage

| PMT Pulse Height | High Voltage Level | |
|---|---|---|
| | LABRVD | E1198-26 |
| 30 mV | -790 V | -600 V |
| 60 mV | -880 V | -670 V |
| 90 mV | -970 V | -740 V |

| Reaction | $E_{res}$ [keV] | γ-ray energy [keV] |
|---|---|---|
| $^{39}$K + p = $^{40}$Ca + γ | 1346.6 | 3904.4, 5736.5 |
| $^{23}$Na + p = $^{24}$Mg + γ | 1318.1 | 1368.6, 11584.8 |
| $^{27}$Al + p = $^{28}$Si + γ | 767.2 | 2838.7, 7706.5 |
| $^{23}$Na + p = $^{24}$Mg + γ | 1416.9 | 2754.0, 8925.2 |
| $^{7}$Li + p = $^{8}$Be + γ | 441 | 17619 |
| $^{11}$B + p = $^{12}$C + γ | 675 | 4438.0, 12137.1 |
| $^{11}$B + p = $^{12}$C + γ | 7250 | 22600 |

| | Analog approach | Digital approach |
|---|---|---|
| a | 400 | 6.3 |
| b | 0.625 | 0.625 |

| | c | $28 \cdot 10^{-6}$ | $35 \cdot 10^{-6}$ |
|---|---|---|---|

| # | Detector Size | PMT | Voltage Divider | HV | CFD T.D. | Int. FWHM [ps] |
|---|---|---|---|---|---|---|
| A | 1"x1" | XP 20 D0B | 184K/T | -500 V | 16 ns | 363 |
| B | 1.5"x1.5" | R6231 | AS20 | +500 V | 16 ns | 646 |
| C | 3"x3" | R6233-100SEL | LABRVD | -500 V | 20 ns | 671 |
| D | 3.5"x8" | R10233-100SEL | LABRVD | -500 V | 20 ns | 880 |